\documentclass[AMA,STIX1COL]{WileyNJD-v2}
\articletype{Article Type}

\received{}
\revised{}
\accepted{}

\usepackage{graphicx}
\usepackage{subfigure}
\usepackage{amsmath}
\usepackage{booktabs}
\usepackage{algorithm}
\usepackage{color}
\usepackage{mathtools}
\usepackage{diagbox} 
\usepackage{algpseudocode}
\usepackage{verbatim}
\usepackage{flushend}
\usepackage{amsthm}
\usepackage{multirow}
\usepackage{algorithmicx}
\usepackage{perpage} 
\MakePerPage{footnote} 

\allowdisplaybreaks 
\algnewcommand\algorithmicinput{\textbf{Input:}}
\algnewcommand\INPUT{\item[\algorithmicinput]}
\algnewcommand\algorithmicoutput{\textbf{Output:}}
\algnewcommand\OUTPUT{\item[\algorithmicoutput]}
\algnewcommand\algorithmicbegin{\textbf{begin}}
\algnewcommand\BEGIN{\item[\algorithmicbegin]}
\algnewcommand\algorithmicendbegin{\textbf{end}}
\algnewcommand\ENDBEGIN{\item[\algorithmicendbegin]}

\algdef{SE}[DOWHILE]{Do}{doWhile}{\algorithmicdo}[1]{\algorithmicwhile\ #1}%

\usepackage{hyperref}

\makeatletter
\def\@fnsymbol#1{\ensuremath{\ifcase#1\or \dagger\or * \or \ddagger\or
   \mathsection\or \mathparagraph\or \|\or **\or \dagger\dagger
   \or \ddagger\ddagger \else\@ctrerr\fi}}
    \makeatother

\newcommand{\rev}[1]{{\color{black}{#1}}}

\def \figwidth {0.48}

\begin{document}

\title{Efficient Asynchronous Federated Learning with Sparsification and Quantization}

\author[1,3]{Juncheng Jia $^\dagger$}
\author[2]{Ji Liu $^\dagger$}
\author[1]{Chendi Zhou}
\author[1]{Hao Tian}
\author[4]{Mianxiong Dong}
\author[5]{Dejing Dou}

\authormark{Tian and Liu \textsc{et al}}

\address[1]{\orgname{School of Computer Science and Technology, Soochow University}, \orgaddress{\state{Suzhou}, \country{China}}}
\address[2]{\orgname{Hithink RoyalFlush Information Network Co., Ltd.}, \orgaddress{\state{Hangzhou}, \country{China}}}
\address[3]{\orgname{Collaborative Innovation Center of Novel Software Technology and Industrialization}, \orgaddress{\state{Suzhou}, \country{China}}}
\address[4]{\orgname{Department of Sciences and Informatics, Muroran Institute of Technology}, \orgaddress{\state{Muroran}, \country{Japan}}}
\address[5]{\orgname{Boston Consulting Group}, \orgaddress{\state{Beijing}, \country{China}}}

\corres{
$\dagger$ Equal contribution.\\
* Ji Liu. \email{jiliuwork@gmail.com}}


\abstract[Summary]{
While data is distributed in multiple edge devices, Federated Learning (FL) is attracting more and more attention to collaboratively train a machine learning model without transferring raw data. FL generally exploits a parameter server and a large number of edge devices during the whole process of the model training, while several devices are selected in each round. However, straggler devices may slow down the training process or even make the system crash during training. Meanwhile, other idle edge devices remain unused. As the bandwidth between the devices and the server is relatively low, the communication of intermediate data becomes a bottleneck. In this paper, we propose Time-Efficient Asynchronous federated learning with Sparsification and Quantization, i.e., TEASQ-Fed. TEASQ-Fed can fully exploit edge devices to asynchronously participate in the training process by actively applying for tasks. We utilize control parameters to choose an appropriate number of parallel edge devices, which simultaneously execute the training tasks. In addition, we introduce a caching mechanism and weighted averaging with respect to model staleness to further improve the accuracy. Furthermore, we propose a sparsification and quantitation approach to compress the intermediate data to accelerate the training. The experimental results reveal that TEASQ-Fed improves the accuracy (up to 16.67\% higher) while accelerating the convergence of model training (up to twice faster).
}


\maketitle

\section{Introduction}
With more and more data generated and distributed in edge devices (devices), e.g., mobile phones and Internet of Things (IoT) devices, a huge amount of data can be exploited to train a deep learning model for diverse artificial intelligent applications  \cite{wang2019adaptive}, e.g., image classification  \cite{rawat2017deep}, keyboard prediction  \cite{hard2018federated}, location-based social networks\rev{\cite{qi2022privacy, liu2023privacy}}, and etc. Centralized training approaches generally aggregate all the decentralized data to a server or a data center for the training process. These approaches may incur significant huge communication overhead and bring severe privacy security risks \cite{yang2019federated}. As an emerging distributed machine learning approach, Federated Learning (FL) \cite{mcmahan2017communication,liu2022distributed} can enable collaborative training of a machine learning model with distributed raw data  while protecting the privacy of the data. 

Within the training process of FL, models are updated with the raw data in edge devices and aggregated in a parameter server (server) \cite{liu2021heterps}. The aggregation of models can be carried out synchronously \cite{mcmahan2017communication,zhou2022efficient} or asynchronously \cite{mohammad2019adaptive}. 
In each round of the whole training process, the server sends the global model to selected devices for local updates before the aggregation of models. The devices update the received model utilizing local data. Then, the devices send back the updated models to the server. With the synchronous method, the server aggregates the updated models to  a new global model after receiving the updated models from all the devices. However, devices may have heterogeneous limited computing resources and communication capabilities, e.g., limited computing capabilities, low battery energy, and low bandwidth \cite{li2020federated}. As a result, the training process may take a long time, and the communication between devices and the server is inefficient. Furthermore, synchronous aggregation cannot fully exploit the edge devices as the unselected devices remain idle during each global epoch. In addition, some powerful devices need to wait for low devices, i.e., stragglers, to continue following updates. In contrast, with the asynchronous aggregation, the server can update the global model immediately once receiving an updated model from any selected device \cite{xie2019asynchronous}, which can take advantage of all the available devices to achieve better performance.

Although asynchronous aggregation can prevent the server from waiting for slow devices, it still suffers from accuracy degradation and inefficient data communication. First, the staleness of the updated models may degrade the accuracy of the global model. For instance, at a given time, the server aggregates the latest models of powerful devices. A show device may just finish the update of a very old global model. Then, the updated model from the slow device may degrade the accuracy of the global model after the aggregation in the server. Second, the issue of non-Independent and Identically Distributed (non-IID) data may also degrade the accuracy of the global model. As the data is generally non-IID, the local optimal model may differ from the global optimal model. The direct aggregation of models from devices may correspond to a global model of inferior accuracy. Finally, as the bandwidth between the devices and the server is limited, it takes much time to transfer the models without sparsification \cite{stich2018sparsified} or quantization \cite{aji2017sparse}.

In this paper, we propose a Time-Efficient Asynchronous FL approach with Sparsification and Quantization (TEASQ-Fed) to improve the training efficiency and the model accuracy of FL. 
We exploit the idle time of devices more aggressively within the training process. While in the existing schemes devices wait for the task assignment from the server passively, in this work we enable idle devices to apply for training tasks actively. However, in order to reduce the slow convergence brought by large numbers of parallel devices participating in the training process, we propose a control parameter \emph{C}-fraction. In addition, we utilize a regularization term to improve the stability of the training process with non-IID data. Moreover, to alleviate the staleness issue of asynchronous training, we exploit a caching mechanism and aggregate models with respect to staleness. Finally, we propose a dynamic decaying approach to choose appropriate compression parameters while exploiting sparsification and quantization to reduce the communication cost during the training process. This paper is an extended version of a conference version \cite{zhou2021tea}. The main contributions of this paper are:

\begin{enumerate}
    \item The design of an asynchronous FL approach, i.e., TEASQ-Fed, to fully exploit devices for the training process. We propose a control parameter to limit the concurrent device participation in order to achieve fast convergence.
    \item A model aggregation approach composed of a regularization method and staleness based mechanism to address the problem of non-IID data and the impact of staleness.
    \item A dynamic decaying approach to generate appropriate compression parameters while exploiting sparsification and quantization for efficient data transfer between devices and the server (extra contribution compared with \cite{zhou2021tea}). 
    \item An extensive evaluation, based on Fashion-MNIST and a convolutional neural network with 100 devices, diverse data distributions, and multiple combinations of control parameters of TEASQ-Fed, compared with baseline approaches (with extra extensive experimental results for the
    sparsification and quantization approach compared with \cite{zhou2021tea}).
\end{enumerate}

The rest of this paper is structured as follows. Section~\ref{sec:relatedwork} reviews the related work. Section~\ref{sec:problem} formulates optimization problems of FL. Section~\ref{sec:approach} provides the detailed explanation of TEASQ-Fed. In Section~\ref{sec:experiment}, we conduct extensive experiments with different data sets and data distributions and demonstrate the impact of hyper-parameters. Finally, Section~\ref{sec:conclusion} concludes this paper.

\section{Related Work}
\label{sec:relatedwork}

With the wide usage of edge devices, the computation is carried out in edge devices, i.e., edge computing, to deal with the large amounts of data generated and distributed \cite{bouadjenek2018distributed} in edge devices. As an emerging machine learning paradigm, FL is an effective solution to address the communication delay, data scalability, and privacy challenges brought by the centralized training process \cite{chen2019deep,Liu2022From}.
FL has been widely used in multiple areas with the integration with edge computing, e.g., keyboard prediction \cite{hard2018federated}, edge intelligence \cite{wang2019edge}, and video surveillance \cite{chen2019distributed}. However, the training process of FL with existing approaches faces two challenges, i.e., accuracy degradation due to non-IID data and low efficiency owing to a huge amount of heterogeneous devices of modest computation and communication capacity. 

While the convergency of FedAvg is theoretically analyzed with non-IID data \cite{li2019convergence}, some other methods are proposed to address the accuracy degradation brought by non-IID data within the synchronous model aggregation mechanism. For instance, FedProx \cite{li2020federated} is proposed based on regulization on each device to address the non-IID data issue. A data sharing method \cite{zhao2018federated} is exploited to improve the accuracy with non-IID data. However, these methods cannot be directly exploited in the asynchronous mechanism. 

Existing FL approaches generally exploit a synchronous model aggregation mechanism, which corresponds to inefficient training. FedAvg \cite{mcmahan2017communication} is a representative synchronous model aggregation mechanism, which randomly selects available devices for the training. To deal with the heterogeneity of devices in terms of communication and computing capabilities, diverse device selection methods are proposed  \cite{fu2023client}, e.g., greedy-based \cite{nishio2019client}, Bayesian optimization-based \cite{zhou2022efficient}, and reinforcement learning-based \cite{zhou2022efficient}. \rev{PyrmidFL\cite{li2022pyramidfl} exploits data and system heterogeneity within selected clients, determines the utility-based client selection and then optimizes utility profiling locally. Although these methods can improve the efficiency with proper device selection, the server still needs to wait until the updated model of the slowest selected device is uploaded to the server. }

\rev{Asynchronous methods \cite{xie2019asynchronous, xu2023asynchronous} are proposed to address the straggler issue and improve the utilization of devices. FedBuff\cite{nguyen2022federated} is an asynchronous federated optimization framework using buffered asynchronous aggregation, where clients conduct local training and communicate with the server asynchronously. The server aggregates multiple client updates in a secure buffer before performing a server update. 
Based on the notion of concurrency, i.e., the number of workers that compute gradients in
parallel, theoretical analysis shows a much faster convergence rate for asynchronous FL \cite{koloskova2022sharper}.
However, in these existing asynchronousy FL methods, the devices are passively triggered by the server for model training and some idle devices are not utilized, meanwhile the advanced data compression for transmission is not incorporated.  
A semi-asynchronous method SAFA is proposed to improve the efficiency of FL \cite{wu2020safa}, which does not consider the non-IID data within diverse devices. Similarly, a layerwise asynchronous model update and temporally weighted aggregation are proposed to improve communication efficiency \cite{chen2019communication}, which still corresponds to synchronous communication in terms of the whole model. 
Asynchronous federated unlearning \cite{su2023asynchronous} is also studied in the literature, which divides the clients into clusters (subsets), conducts independent and asynchronous model training for clusters (subsets). The unlearning cost is limited due to the client partitioning. Federated unlearning is out of the scope of this paper.}

In order to address the communication bottleneck with FL, diverse techniques \cite{che2023federated,liu2023distributed,che2023fast,che2022federated,liu2022multi,Li2022FedHiSyn,Liu2024AEDFL,Li2024On,Liu2024FedASMU,Jin2022Accelerated} can be exploited, e.g., sparsification, quantization, and pruning. Model sparsification transfer has been widely exploited in distributed machine learning  \cite{xu2021grace,aji2017sparse,xie2020cser, stich2018sparsified,basu2019qsparse}. A typical sparsification method is the Top-K sparsification \cite{stich2018sparsified}, where the K largest values in the tensor are selected and retained, and the rest of the values are set to zero so that the sparse tensor will be easier to compress and transmit in practical application scenarios. Model quantization is also utilized in distributed machine learning as well as FL. Quantization is the process of compressing the representation of values with a smaller number of bits to reduce the size of the values \cite{aji2017sparse}. The reset transmission error method is proposed to remend the global model by adjusting the errors brought by the quantization or sparsification \cite{xie2020cser}. However, this method incurs extra synchronizations and leads to significant communication overhead. In addition, model pruning techniques \cite{jiang2019model, Zhang2022FedDUAP, zhang2021validating} can be exploited to reduce the huge overhead of computation and communication during the whole training process. \rev{In FedCG\cite{jiang2023heterogeneity}, the server selects a representative client subset for local training considering statistical heterogeneity and then these selected clients upload compressed model updates matching their capabilities for aggregation, which alleviates the communication load and mitigates the straggler effect. FedLamp \cite{xu2022adaptive} adaptively determines diverse and appropriate local updating frequencies and model compression ratios in the resource-constrained edge computing systems, so as to reduce the waiting time and enhance the training efficiency.
However, these above approaches cannot choose the appropriate parameters for the data compression of both sparsification and quantization with accuracy degradation constraints, which may lead to unexpected accuracy degradation.}

In this paper, we propose a new time-efficient approach to efficiently exploit devices to train a global model in FL while addressing the problem brought by non-IID data. In addition, we exploit sparsification and quantization with proper data compression parameters to improve communication efficiency while satisfying the accuracy degradation constraints.

\section{Problem Formulation}
\label{sec:problem}

\begin{table} [!t]
\caption{Summary of Main Notations}
\label{table1}
\begin{center}
\begin{tabular}{cc}
\hline
Notation & Definition \\
\hline

$N$; $k$ &  The number of devices; the index of devices \\
$\mathcal{D}_k$; $n_k$ & Local dataset on device $k$; size of samples \\
$w^g$; $S_{w^g}$ & The global weight; size of the global weight \\
$w^l$; $S_{w^l}$ & The local weight; size of the local weight \\
$p_s$; $p_q$ & Sparsification parameter; quantization parameter \\
$r^d_k$ & The maximum transmission rate of downloading global models for device $k$ \\
$r^u_k$ & The maximum transmission rate of uploading local models for device $k$ \\
$S_{w^g, p_s, p_q}$ & The size of compressed global model weights \\
$S_{w^l, p_s, p_q}$ & The size of compressed local model weights \\
$L^{down}_{k,t}$ & The latency of downloading global model weights \\
$L^{up}_{k,t}$ & The latency of uploading global model weights \\
$L^{cp}_{k,t}$ & The latency of  model computation \\
$L^{round}_{k,t}$ & The total time of round $t$ for device $k$ \\
$L_T$ & The total time for $T$ rounds of asynchronous FL \\
$Set_p$ & The set of candidate parameter $p_s$ \\
$Set_q$ &  The set of an available number of bits in the quantization process $p_q$ \\

\hline
\end{tabular}
\end{center}
\end{table}

In this section, we formulate the problem to address in this paper. We first present the overview of an FL system. Then, we present the latency model. Finally, we present the problem statement. The summary of the main notations is shown in Table \ref{table1}.

We consider an FL system consisting of a server module (server) and \emph{N} devices. We consider idle time as the time slot when a device has no task to execute. Once receiving an updated model from a device, the server aggregates the global model with the updated model. $n_k$ samples are stored on the $k$-th device where $k\in\{1,...,N\}$. The total number of samples can be calculated as $n =\sum_{k=1}^Nn_k$. The training objective is to update the global model with the weight parameters $w$ utilizing the local data of all edge devices, which is formulated as follows:
\begin{equation}
  \min_{w \in \mathbb{R}^d} f(w)=\sum_{k=1}^{N}\frac{n_k}{n}{\rm E}_{x_k \sim \mathcal{D}_k}[f_k(w;x_k)]\label{eq:eq11}
\end{equation}
where $x_k$ is sampled from the local dataset $\mathcal{D}_k$ on device $k$, and $n_k=|\mathcal{D}_k|$. $f_k(w;x_k)=l(w;x_k)$ is the loss calculated on device $k$ with the sample $x_k$ and the parameters $w$. We assume that the data distributions on devices are not Identically Independently Distributed, i.e., non-IID.

\subsection{Latency Model}

In this section, we present the model to calculate the latency for the training process, including communication latency, computation latency, and round latency.

\subsubsection{Communication Latency}

When the server and devices communicate with each other, either the server sends the global model to the devices or the devices upload updated local models to the server. We denote the global model weight as $w^g$, which consists of values in float32 without quantization, and we denote the size of $w^g$ as $S_{w^g}$ in bits. Similarly, the size of local model weight $w^l$ is denoted as $S_{w^l}$ in bits. When exploiting sparsification and quantization with the corresponding parameters of $p_s$ and $p_q$, respectively, the size of local model weight becomes $S_{w^l, p_s, p_q}$. $p_s$ represents the ratio of the non-zero values kept in the model within the sparsification process and $p_q$ corresponds to the number of bits, e.g., float32 or int8, to represent each value within the quantization process. The maximum transmission rate of the $k$-th device (bits/s), where $k\in\{1,...,N\}$, for the server to send global models is defined as $r^d_k$. Similarly, the maximum transmission rate of devices uploading local models (bits/s) is defined as $r^u_k, \forall k \in \{1,...,N\}$. We assume that $r^u_k$ and $r^d_k$ remain constant during training. Thus, for each device $k \in\{1,...,N\}$ and each round $t \in\{1,...,T\}$, the global model download latency is $L^{down}_{k,t} = \frac{S_{w^g, p_{s}, p_{q}}}{r^d_k}$, and the local model upload latency is $L^{up}_{k,t} = \frac{S_{w^l, p_{s}, p_{q}}}{r^u_k}$.

\subsubsection{Computation Latency}

To characterize the randomness of the computation latency of local model update, we exploit the shifted exponential distribution \cite{shi2020joint}:
\begin{equation}\label{eq:5}
    \mathbb{P}[L^{cp}_{k,t} < l] = \begin{cases}
                                   1 - e^{- \frac{\phi_k}{\tau b}(l-a_k\tau b)} , &  l \ge a_k \tau b, \\
                                   0 , & \rm{otherwise},
                                   \end{cases}
\end{equation}
where $a_k>0$ represents the maximum of computation capabilities, and $\phi_k>0$ represents the fluctuation  of computation capabilities. We assume that $a_k$ and $\phi_k$ remain constant during training. The computational delay of device $k$ in round $t$ is denoted as $L^{cp}_{k,t}$.
 The latency of model computation (i.e., $L^{cp}_{k,t}$) can also be affected by the time to compress the local model and the time to decompress the global model, but since these two components of latency are usually much smaller than the time to update the model, they are ignored in this paper.

\subsubsection{Round Latency}

For a certain round $t$, and a certain device $k$, the total time of a round, i.e., the time period between the downloading of the global model to the uploading of the trained local model, is defined as:
\begin{equation}
L^{round}_{k,t} = L^{down}_{k,t} + L^{cp}_{k,t} + L^{up}_{k,t},
\end{equation}
when device $k$ uploads its updated local mode to the server in round $t$.

We define the total time for $T$ rounds of asynchronous FL as $L_T$. However, due to the asynchronous nature among multiple rounds, we cannot simply sum them up to obtain the total time consumed by the $T$-round asynchronous FL. In addition, because of asynchronicity, we cannot intuitively describe the relationship between $L_T$ and $L^{round}_{k,t}, \forall{ t \in \{ 1,...T\} }$ in numerical terms.  $L_T$ depends on $L^{round}_{k,t}, \forall{t \in \{ 1,...T\}}$, and a smaller $L^{down}_{k,t}$, $L^{cp}_{k,t}$, or $L^{up}_{k,t} $ can reduce $L_T$ in the asynchronous FL system. Then, we describe the relationship by a function \rev{$g(\cdot)$:
\begin{equation}
    L_T = g(L^{round}_{k_1,1}, L^{round}_{k_2,2}, ..., L^{round}_{k_T,T}),
\end{equation}
where $k_t \in \{1, 2, ..., N\}$ represents the device for round $t$, i.e., each round is indicated by the update of device $k_t$. $g(\cdot)$ represents the total time of the whole asynchronous FL process.}

\subsection{Problem Statement}

We formulate the overall optimization objective as follows:
\begin{align*}
   \min_{w} f(w)& =\sum\nolimits_{k=1}^{N}\frac{n_k}{n}{\rm E}_{x_k \sim \mathcal{D}_k}[f_k(w, p_s, p_q;x_k)]\label{eq:eq1}\\ 
    s.t. \qquad & g(p_s, p_q; L^{round}_{k_1,1}, L^{round}_{k_2,2}, ..., L^{round}_{k_T,T})  \leq L,   \\
    & k \in \{ 1,... N \},   \\
    & t \in \{ 1,...T \}, \\
    & p_s \in Set_{p}, \\
    & p_q \in Set_{q}, \\
    & w \in \mathbb{R}^d
\end{align*}
where $x_k$, $n_k=|\mathcal{D}_k|$, $f_k(\cdot)$ are the same as those in Formula~\eqref{eq:eq11}. $L$ is the total time budget and $g(\cdot)$ is the estimated training time. $p_s$ is the parameter for sparsification and $p_q$ is the parameter for quantization. 
$Set_p$ represents the set of candidate parameters $p_s$, $Set_{q}$ represents the set of an available number of bits in the quantization process.

\section{TEASQ-Fed Protocol}
\label{sec:approach}

We propose a new asynchronous FL approach, i.e., TEASQ-Fed, to collaboratively train a model. TEASQ-Fed consists of four parts. First, we asynchronously exploit the edge devices in the training process of FL. Second, we utilize a regularization method to address the problem of non-IID data and the impact of staleness. Third, we generate appropriate compression parameters while exploiting sparsification and quantization. In this section, we present the details of TEASQ-Fed, including the protocol overview, task management, model aggregation, sparsification, and quantization.

\subsection{Asynchronous Protocol Overview}

\begin{figure}[!t]
\centerline{\includegraphics[width=\linewidth]{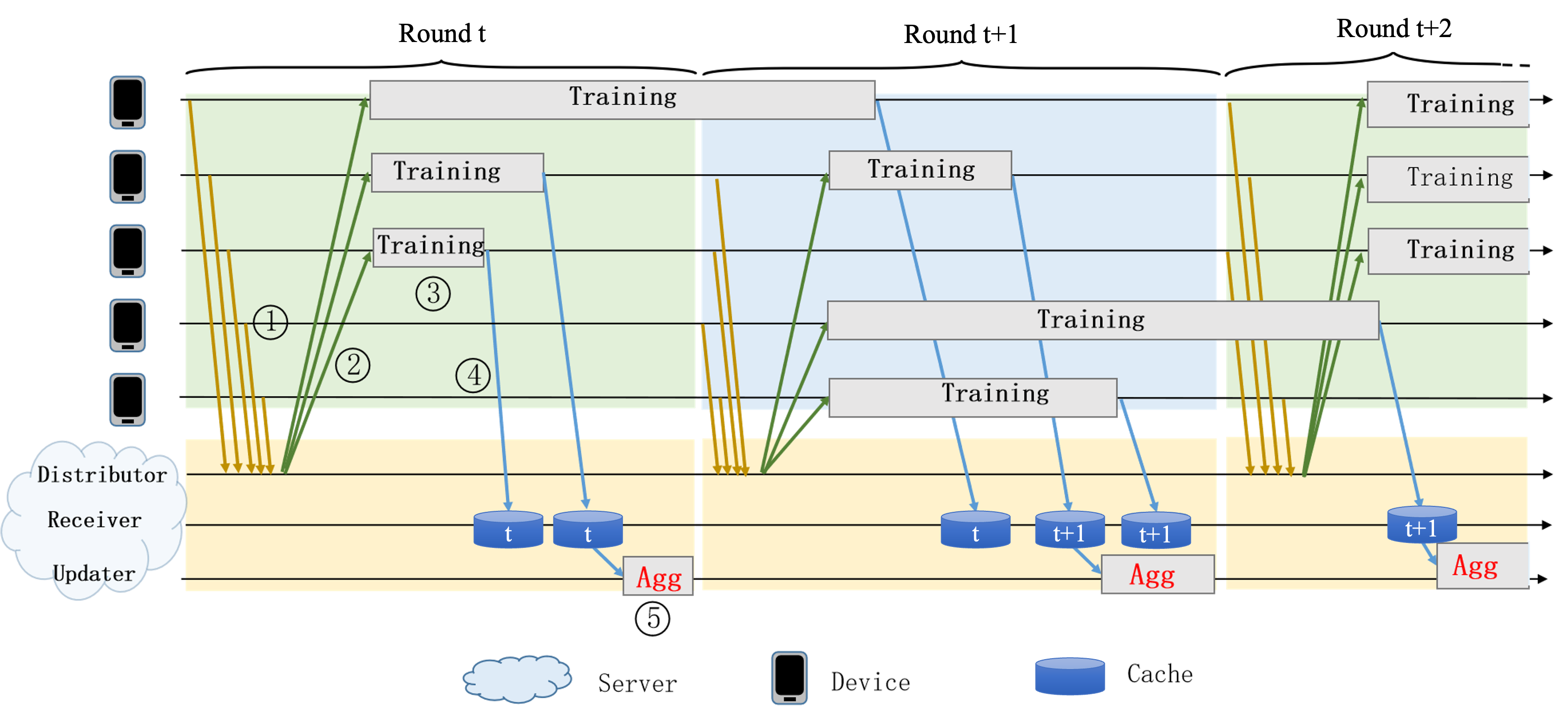}}
\caption{\rev{The overview of TEASQ-Fed protocol.}}
\label{fig:framework}
\end{figure}

Diverse devices are exploited based on their private local data in the training process of FL, while the data distribution is generally heterogeneous, i.e., non-IID. Moreover, in the synchronous training method (e.g., FedAvg \cite{mcmahan2017communication}), the selected devices may be busy or crash while performing training tasks, which may prolong the training process. On the contrary, in our proposed asynchronous protocol edge devices initiate training task requests proactively when they are idle, which can alleviate the device unavailability issue during training.
Besides, we propose a cache mechanism and a weighted averaging method with regularization for the server to increase the convergence speed of the model training.

\rev{The overview of TEASQ-Fed protocol is shown in Figure~\ref{fig:framework}. At the beginning, the server initializes a global model randomly. The idle edge device sends a request to the server to obtain a training task (Step \textcircled{\footnotesize 1} in Figure~\ref{fig:framework}). 
After receiving the task request from the device, the server checks whether the number of devices currently participating in the latest model training is less than a limit. If the limit is not reached, the latest global model is delivered to the device and change the record of the number of devices participating in training the latest global model, otherwise it is not delivered (Step \textcircled{\footnotesize 2}).
The idle devices receiving the global model update the global model asynchronously using their local data and upload the updated model parameters to the server (Step \textcircled{\footnotesize 3}).
The server puts the updated model and time stamp received from devices into the cache, and change the record of the number of these devices participating in training in the same stamp (Step \textcircled{\footnotesize 4}). The server uses weighted averaging according to staleness to update the global model after it receives updates from a certain number of devices, and then enters the next round of global training (Step \textcircled{\footnotesize 5}).
All steps \textcircled{\footnotesize 1} - \textcircled{\footnotesize 5} are iterated until the global model achieves the desired performance.}

\rev{The detailed design of TEASQ-Fed protocol is explained in the following subsections. In Subsection \ref{subsec:task} and \ref{subsec:local update}, we describe the training task request and local model update (Algorithm \ref{alg:Alg1_new}). Idle devices request training tasks from the server proactively. When the number of parallel devices participating in the training process is significant, the final convergence may degrade due to the diverse version of the original model. Thus, we set a limit on the number of parallel training devices when distributing training tasks. In Subsection \ref{subsec:aggregation}, we describe the global model update on the server (Algorithm \ref{alg:Alg2_new}). We adopt a caching mechanism with a weighted average mechanism for model aggregation to further improve the accuracy. In Subsection \ref{subsec:compress}, we describe the dynamic weight compression method with sparsification and quantization as well as compression parameter selection (Algorithm \ref{alg:compress}, \ref{alg:decompress} and \ref{alg:search}).}

\begin{algorithm}[!t]	
    \caption{Training task request and local model training.}
    \label{alg:Alg1_new}
        \begin{algorithmic}
        \renewcommand{\algorithmicrequire}{ \textbf{Input:}}  
        \renewcommand{\algorithmicensure}{ \textbf{Output:}}   
        \Require{\\
            $N$: the quantity of edge devices\\
            $T$: the quantity of rounds \\
            $E$: the quantity of local updates per round \\
            $B$: minibatch size of local updates \\
            $\eta$: learning rate \\ 
            $\mu$: the weight of regularization \\
            $P$: the quantity of edge devices taking part in training during the $t$-th training round \\
            $C\in\left(0,1\right)$: the proportion of edge devices
         }
    \end{algorithmic}
    {\bf{Server process:}} \hspace{0.3in}{// on the server}\\
    {Initialize $w^0$, $t\leftarrow0$, $P\leftarrow0$} \\ 
    \hspace{-2.6in}{\emph{Distributor:}}
    \begin{algorithmic}[1]
        \State Receive request from the inactive edge device
        \If{$P \leq [N\cdot C]$} \hspace{0.3in}{//$P$ represents the number of selected devices} 
        \label{alg:verify}
            \State Compress the latest model $w^t_{compress} = \mathcal{C}(w^t, p_s, p_q)$
            \State Transfer the latest compressed model $w^t_{compress}$ to the requested device
            \State Update $P\leftarrow P+1$
        \EndIf
   \end{algorithmic}
   {\bf{Device process:}} \hspace*{0.3in}{// on inactive device $k$}
   \begin{algorithmic}[1]
        \If{device $k\in \left [N\right]$ is idle}
            \State Send training task request to the server
            \If{receive $\left (w^t_{compress},t\right)$ from the server}
                \State Decompress  $w^t = \mathcal{C}^{-1}(w^t_{compress})$
                \State $\chi_k\leftarrow$ (split $\mathcal{D}_k$ into batches of size $B$) \State $w^{h_k}_k \leftarrow w^t$
                \For{each local epoch $i$ from 1 to $E$}
                    \For{batch $x_k\in \chi_k$}
                        \State $w^{h}_k= w^{h}_k-\eta \left(\nabla f_k(w^{h}_k;x_k)+\mu(w^{h}_k-w^t)\right)$
                    \EndFor
                \EndFor
                \State Compress the updated local model $w^{h}_{compress, k} = \mathcal{C}(w^{h}_k, p_s, p_q)$
                \State Transfer the updated local model and time stamp $ (w^{h}_{compress, k},h)$ to the server 
            \EndIf
        \EndIf
   \end{algorithmic}
\end{algorithm}

\subsection{Task Management}
\label{subsec:task}

\begin{algorithm}[!t]
    \caption{Model aggregation.}
    \label{alg:Alg2_new}
    \begin{algorithmic}
        \renewcommand{\algorithmicrequire}{ \textbf{Input:}}  
        \renewcommand{\algorithmicensure}{ \textbf{Output:}}   
        \Require{\\
            $N$: the quantity of edge devices\\
            $T$: the quantity of rounds \\
            $\gamma \in\left(0,1\right)$: cache fraction \\
            $\alpha \in\left(0,1\right]$: the hyper-parameter for aggregation
         }
    \end{algorithmic}
    {\bf{Server process:}} \hspace*{0.1in}{// performing on the server}\\
    \emph{Receiver:}
    \begin{algorithmic}[1]
        \State Get an update $w^{h_c}_{compress, k_c}$ and the time stamp $h_c$ from any inactive device
        \State Push $\left( w^{h_c}_{compress, k_c} \right)$ into the cache queue $Q$
        \State Update $P\leftarrow P-1$
   \end{algorithmic}
   \emph{Updater:}
   \begin{algorithmic}[1]
        \For{round $t =0,1, ..., T-1$}
            \For{$c = 1$ to $\left[N\cdot\gamma\right]$}
                \State Pop $\left( w^{h_c}_{compress, k_c} \right)$ from the cache queue $Q$
                \State Decompress $w^{h_c}_{k_c} = \mathcal{C}^{-1}(w^{h_c}_{compress, k_c})$
            \EndFor
            \State Calculate the aggregated model weight according to Formula \eqref{eq:eq4} and \eqref{eq:eq5}
            \State Calculate the global model $w^{t+1}$ according to Formula \eqref{eq:eq6} and \eqref{eq:eq7}
        \EndFor
   \end{algorithmic}
\end{algorithm}

\rev{With existing FL systems \cite{mcmahan2017communication,nishio2019client}}, the devices are generally randomly selected by the server in each round. Insufficient battery power, unstable network conditions, and device crashes may affect the training of some devices, making them unreliable. Moreover, after completing the training task, the selected device may remain idle. The device will not be selected again until the next round of training. Furthermore, idle devices other than the selected one are not fully utilized. As a strategy for improving the utilization of devices, we exploit asynchronous communication to enable idle devices to actively participate in training tasks. Thus, if the device is assigned a training task, it sends a task request to the server once it becomes idle, and the server sends the latest model to it.

As large amounts of device participation within the FL training process may incur slow convergence, 
The hyper-parameter $C$ controls how many devices participate in the parallel training from the same global model. By implementing the $C$-fraction, the model \cite{li2019convergence} can be guaranteed to have a speedy convergence and prevent server overload due to excessive participants. Algorithm~\ref{alg:Alg1_new} indicates that when a task request comes from a device that is idle, the server determines whether to immediately serve the latest model (Line~\ref{alg:verify}). Devices taking part in the latest model training will receive the latest model only if the number is less than $\left[N \cdot C\right]$; otherwise, it will not. A model received from the server will then be updated by the idle device with local data once it receives a model from the server. 
The TEASQ-Fed suffers from a slow convergence speed, if $C$ is set too small. A large $C$ will increase the risk of congestion and also slow down the convergence speed. It is true that the idle devices may introduce biases; however, we believe that these biases are negligible since our approach provides a much higher degree of accuracy than FedAvg (see details in Section~\ref{subsection:results}).

As shown in Algorightm~\ref{alg:compress}, $\mathcal{C}(w, p_s, p_q)$ compresses all the tensors in model $w$ to get compressed model weights $w_{compressed}$, i.e., $w_{compressed}$ = $\mathcal{C}(w, p_s, p_q)$. The sparsification \cite{stich2018sparsified} operation is explained in Lines~\ref{al:sparbegin}-\ref{al:sparend}. The quantization \cite{alistarh2017qsgd} operation is explained in \rev{Line \ref{al:quan}}. We  remove these zero elements in Line~\ref{al:removeZero}. 

\subsection{Local Update}
\label{subsec:local update}

Since diverse devices store data in non-IID manners \cite{li2020federated}, the local updates of some devices may slow down the global model convergence speed.
We exploit a penalty term based on \cite{li2020federated, xie2019asynchronous} to reduce adverse impacts of the non-IID data. The loss function in an idle device $k$ contains a regularization term with the global model $w$ received from the server as defined:
\begin{equation}
  \min \limits_{w \in \mathbb{R}^d}{\rm E}_{x_k \sim \mathcal{D}_k}[f_k(w;x_k)]+\frac{\mu}{2}\parallel w-w^t\parallel ^2\label{eq:eq2}
\end{equation}
Regularization weights are parameterized as $\mu$. Using the penalty term can reduce data heterogeneity and make the model more stable.

\subsection{Model Aggregation}
\label{subsec:aggregation}


The server and idle devices are updated asynchronously during model training, as shown in Algorithm~\ref{alg:Alg2_new}. The server puts the received updates into the cache and utilizes them for aggregation later. With the help of the hyper-parameter $\gamma$, we limit the size of the cache or the total number of uploaded updates in the cache. We first execute elastic averaging between the most recent local averaging update and the current global model, which not only utilizes the knowledge of new updates but also does weighted averaging on the cached updates. The global model can converge quickly and smoothly in this way.

\subsubsection{Update Caching Mechanism} 

\rev{In the existing asynchronous methods \cite{xie2019asynchronous}, the server immediately updates the global model upon receiving a local update from any device and then moves on to the subsequent training phase.} Although this mechanism results in a significant throughput, a single poor local update could lead to a divergence in the global model. We exploit a caching mechanism to reduce the negative effects of model staleness and increase the stableness of the global model. We set the cache capacity to the size of $K= \left[N \cdot \gamma\right]$ local updates, where $\gamma\in\left(0,1\right)$. We denote the cached updates as $\mathcal{U} = \left\{\left(k_c, w_{k_c}^{h_c}\right)\right\}_{ c \in [0, K]}$, where $w_{k_c}^{h_c}$ is the $c$-th update in the cache from device $k_c$ with time stamp of global model $h_c$. After receiving $K$ local updates, the server changes the global model through weighted averaging depending on the staleness of each update. In the paper, we set the value of $\gamma$ to 0.1.

\begin{algorithm}[!t]		
    \caption{Compress the model weight tensor.}
    \label{alg:compress}
    \begin{algorithmic}[1]
        \INPUT{$p_s$, $p_q$, Uncompressed model weight $w$}
        \OUTPUT{Compressed model weight $w_{compressed}$}
        \For{$tensor \in w$}
            \State $values$ $\leftarrow$ Top $p_s\%$ elements of $tensor$ \label{al:sparbegin}
            \State Elements of $tensor$ not in $values$ are set to 0 \label{al:sparend}
            \State Quantize $tensor$ to $p_q$ bits \label{al:quan}
            \State $values, indices \leftarrow$ non-zero elements in $tensor$  \label{al:removeZero}
            \State $tensor \leftarrow concat(values, indices)$ \label{al:concat}
        \EndFor
        \State Combine $tensor$ to $w_{compressed}$  and return
   \end{algorithmic}
\end{algorithm}

\begin{algorithm}[!t]		
    \caption{Decompress the model weight tensor.}
    \label{alg:decompress}

    \begin{algorithmic}[1]
        \INPUT{Compressed model weight $w_{compressed}$}
        \OUTPUT{Decompressed model weight $w_{decompressed}$}
        \For{$tensor_{compressed} \in w_{compressed}$}
            \State $values, indices$ = $tensor_{compressed}$ \label{al:retrieve}
            \State Restore $tensor$ with $values$ and $indices$ \label{al:restoretensor}
            \State Cast $tensor$ to $32$-bits (float32) \label{al:convert}
        \EndFor
        \State Combine $tensor$ to $w_{decompressed}$  and return
   \end{algorithmic}
\end{algorithm}

\begin{algorithm}[!t]	
    \caption{Dynamic data compression.}
    \label{alg:search}
    \begin{algorithmic}[0]
        \INPUT{ \\ 
            $w$: pre-trained model weight \\ 
            $\theta$: the predefined threshold \\
            $Set_s$: the set of sparsification parameters \\
            $Set_q$: the set of quantization parameters \\
            $T$: the maximum number of rounds 
        }
        \OUTPUT{ \\
            $p_{s,t}$ and $p_{q,t}$ $\forall t \in \{1,2, ...,T\}$
        }
    \end{algorithmic}
    \begin{algorithmic}[1]
        \State $acc \leftarrow test(w)$ 
        \State {$p_s$ $\leftarrow$ $p_s' \in Set_s$ with the smallest compression rate}  
        \State {$p_q$ $\leftarrow$ 0}\Comment{No quantization}
        \While{$test(\mathcal{C}^{-1}(\mathcal{C}(w, p_s, p_q))) \geq acc$ - $\theta$ and the compression rate can be reduced} \label{al:searchbegin}
            \While{$test(\mathcal{C}^{-1}(\mathcal{C}(w, p_s)))$ $\geq$ $acc$ - $\theta$}    \label{al:decreasepsbegin}
                \State {$p_s$ $\leftarrow$ $p_s' \in Set_s$ with a bigger compression rate}
            \EndWhile         \label{al:decreasepsend}
            \State $p_q$ $\leftarrow$ $p_q' \in Set_q$ with a bigger compression rate  \label{al:increasepq}
            \While{ $test(\mathcal{C}^{-1}(\mathcal{C}(w, p_s, p_q)))$ $\leq$ $acc$ - $\theta$ }   \label{al:increasepsbegin}
                \State {$p_s$ $\leftarrow$ $p_s' \in Set_s$ with a smaller compression rate}
            \EndWhile       \label{al:increasepsend}
        \EndWhile             \label{al:searchend}
        \State $p_{s, 0}$ $\leftarrow$ the element in $Set_s$ corresponding to a bigger compression rate than that of $p_s$ \label{al:ps0}
        \State $p_{q, 0}$ $\leftarrow$ the element in $Set_q$ corresponding to a bigger compression rate than that of $p_q$ \label{al:pq0}
        \For{$t \in [T]$ }  \label{al:decaystart}
            \State $p_{s,t} \leftarrow$  $p_{s, 0}$ decay $\lfloor \frac{t}{step\_size} \rfloor$ steps
            \State $p_{q,t} \leftarrow$  $p_{q, 0}$ decay $\lfloor \frac{t}{step\_size} \rfloor$ steps
        \EndFor             \label{al:decayend}
        \State Return $p_{s,t}$ and $p_{q,t}$ $\forall t \in [T]$
  \end{algorithmic}
\end{algorithm}

\subsubsection{Staleness-Based Weighted Averaging}

The staleness of local updates in the cache may differ as the models in devices and the server are updated asynchronously. 
We denote the latest global model in round $t$ on server as $w^t$.
Then the staleness of the $c$-th update in the cache $w_{k_c}^{h_c}$ is $t-h_c$.
We define a function $S(\cdot)$ with respect to the staleness as follows:
\begin{equation}
  S\left(t-h_c\right)=\left(t-h_c+1\right)^{-a} , a > 0, \label{eq:eq3}
\end{equation}
where $a$ is a hyper-parameter. After the server caches $K = \left[N \cdot \gamma\right]$ updates, the average update $u$ can be calculated as follows:
\begin{equation}
  u=\frac{\sum\nolimits_{c=1}^{K}S\left(t-h_c\right) \frac{n_{k_c}}{n}w^{h_c}_{k_c}}{\sum\nolimits_{c=1}^{K}S\left(t-h_c\right)\frac{n_{k_c}}{n}}
  =\frac{\sum\nolimits_{c=1}^{K}S\left(t-h_c\right) n_c w^{h_c}_{k_c}}{\sum\nolimits_{c=1}^{K}S\left(t-h_c\right)n_{k_c}},
\label{eq:eq4}
\end{equation}
We use $\delta$ to represent the average staleness of all local weights in the cache,
\begin{equation}
  \delta=\frac{1}{K}{\sum\nolimits_{c=1}^{K}(t-h_c)}.\label{eq:eq5}
\end{equation}
With Formulas \eqref{eq:eq3} and \eqref{eq:eq4}, the value of mixing weight $\alpha^t$ in the $t$-th round is calculate by
\begin{equation}
  \alpha^t=\alpha\cdot S\left(\delta\right),\label{eq:eq6}
\end{equation}
where $\alpha\in\left(0,1\right]$ is the mixing hyper-parameter.
Finally, the server updates the global model $w^t$ with weighted averaging to calculate a new global model $w^{t+1}$ according to
 \begin{equation}
  w^{t+1}=\alpha^t u+\left(1-\alpha^t\right)w^t.\label{eq:eq7}
\end{equation}

\subsection{Dynamic Weight Compression}
\label{subsec:compress}

In order to improve the communication efficiency within the FL training process, we exploit sparsification \cite{aji2017sparse} and quantization \cite{alistarh2017qsgd} to compress the model.
Sparsification compresses a model taking a subset of the original model parameters to represent the original model, e.g., Random-K \cite{stich2018sparsified} and Top-K \cite{aji2017sparse}. Quantization reduces the number of bits in each element of the model parameters from the original 32-bit floating point number to a smaller number of bits, e.g., 8-bit.
We can exploit existing sparsification \cite{stich2018sparsified, aji2017sparse} and quantization \cite{alistarh2017qsgd, basu2019qsparse} compression mechanisms to compress the model.

In Algorithm~\ref{alg:Alg1_new}, $\mathcal{C}(w, p_s, p_q)$ represents the data compression with two steps, i.e., sparsification with parameter $p_s$, and quantization with parameter $p_q$. The decompression operation (i.e., $\mathcal{C}^{-1}(w)$) is the inverse process of the above two steps. The compression method $\mathcal{C}(w, p_s, p_q)$ adapts the Top-K method \cite{aji2017sparse} and the QSGD method \cite{basu2019qsparse} as shown in Algorithm~\ref{alg:compress}. 
The decompression of weights (i.e., $\mathcal{C}^{-1}(w)$) is presented in Algorithm~\ref{alg:decompress}. First, the values and indices are retrieved from the tensor in Line~\ref{al:retrieve}. Then, the original shape of the tensor is restored in Line~\ref{al:restoretensor}. Finally, the type of the tensor is converted back to float32 in Line~\ref{al:convert}.



\begin{figure}[!t]
\centerline{\includegraphics[width=\figwidth\linewidth]{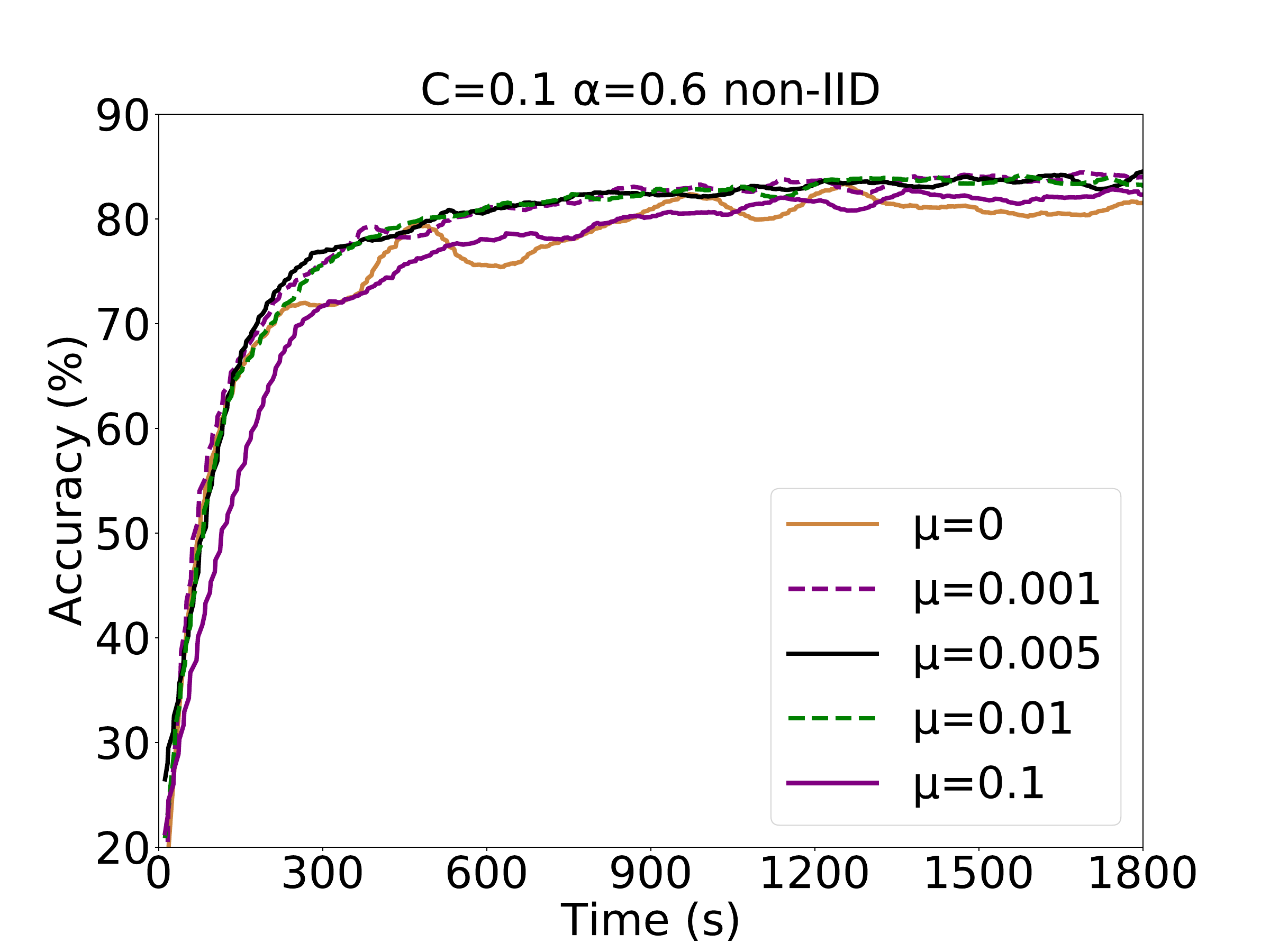}}
\caption{Impacts of $\mu$ in terms of accuracy vs. training time with non-IID dataset.}
\label{fig4}
\end{figure}

With inappropriate values of $p_s$ or $p_q$, the accuracy may be significantly reduced and the data communication efficiency remains low. As shown in Algorithm~\ref{alg:search}, we propose a heuristic method, i.e., dynamic data compression, to generate proper values as $p_s$ and $p_q$ with a decay mechanism. We propose a greedy method to calculate appropriate values for $p_s$ and $p_q$. In order to analyze the impact of the values of $p_s$ and $p_q$ on the accuracy, we take a trained model $w$ to profile the accuracy with diverse values of $p_s$ and $p_q$. First, in Lines~\ref{al:decreasepsbegin}-\ref{al:decreasepsend}, we choose the highest compression rate of $p_s$ without quantization. Then, we take a value of $p_q$ that corresponds to a higher compression rate of quantization in Line~\ref{al:increasepq} and decrease the compression rate of $p_s$ in Lines~\ref{al:decreasepsbegin}-\ref{al:decreasepsend} until the accuracy degradation satisfies a predefined threshold. Afterward, we repeat the aforementioned step of adjustment of $p_s$ in Lines~\ref{al:searchbegin}-\ref{al:searchend}. If the compression can be further reduced, we continue. Otherwise, we finish the search process. In addition, in order to achieve high accuracy with sparsification and quantization, we propose a dynamic method by decaying the compression ratio during the training process in Lines~\ref{al:ps0}-\ref{al:decayend}. 
$\mathcal{C}(w, p_s)$ in Line~\ref{al:decreasepsbegin} represents compressing $w$ with only the sparsification method, and $\mathcal{C}(w, p_s, p_q)$ in Line~\ref{al:increasepsbegin} represents compressing $w$ with both sparsifaication and quantization methods. Lines~\ref{al:ps0}-\ref{al:decayend} represent compression parameter decay process. Based on the $p_s$ and $p_q$ searched, we start with a relatively low compression rate (i.e., $p_{s, 0}$ and $p_{q, 0}$) and increase the compression rate using a constant step size. After determining the compression parameters for each round, the server and each device will compress and decompress model weights in round $t$ using $p_{s,t}$ and $p_{q,t}$.

\section{Experimental Evaluation}
\label{sec:experiment}

We carry out extensive experimentation to evaluate TEASQ-Fed. To demonstrate the efficacy of TEASQ-Fed, we measure the convergence with various machine learning tasks and data distributions.

\subsection{Experiment Setup}

\begin{table}[!t]
\centering
\caption{\rev{The summary of the Fashion-MNIST dataset.}}
\label{table2}
\begin{tabular}{|c|c|}
\hline
Image scale              & 28 $\times$ 28 \\
\hline
Image type              & grayscale  \\
\hline
\# training samples              & 60,000 \\
\hline
\# testing samples             & 10,000 \\
\hline
\# categories             & 10 \\
\hline

\end{tabular}
\end{table}

To evaluate the performance of TEASQ-Fed, we carry out extensive experimentation on the Fashion-MNIST dataset \cite{xiao2017fashion}. As shown in Table \ref{table2}, this dataset includes 70,000 grayscale photos in total, with 60,000 training images and 10,000 testing images, of diverse products from 10 categories. We exploit the dataset to conduct picture categorization with a convolutional neural network (CNN). The CNN consists of two $2\times2$ convolutional layers, a fully connected layer, and a softmax output. We assume that the resource-constrained edge devices are capable of training this lightweight CNN network.

We uniformly distribute the training data among 100 edge devices. We employ an experimental set-up that is similar to the existing work \cite{zhao2018federated} in order to evaluate the performance of TEASQ-Fed under various data distributions. Each device randomly selects a predetermined number of photos from each training batch for the IID setting. The training data is sorted into classes based on categories for non-IID settings. Then, a subset of 2 classes is chosen at random from a total of 10 classes, and each device randomly samples images from this subset. We set $a$ = 0.5 in Formula \eqref{eq:eq3} \cite{xie2019asynchronous} and $\gamma$ = 0.1 in each cycle of the TEASQ-Fed protocol. 

\begin{figure*}[t]
\centering
\subfigure[Non-IID]{\includegraphics[width=\figwidth\linewidth]{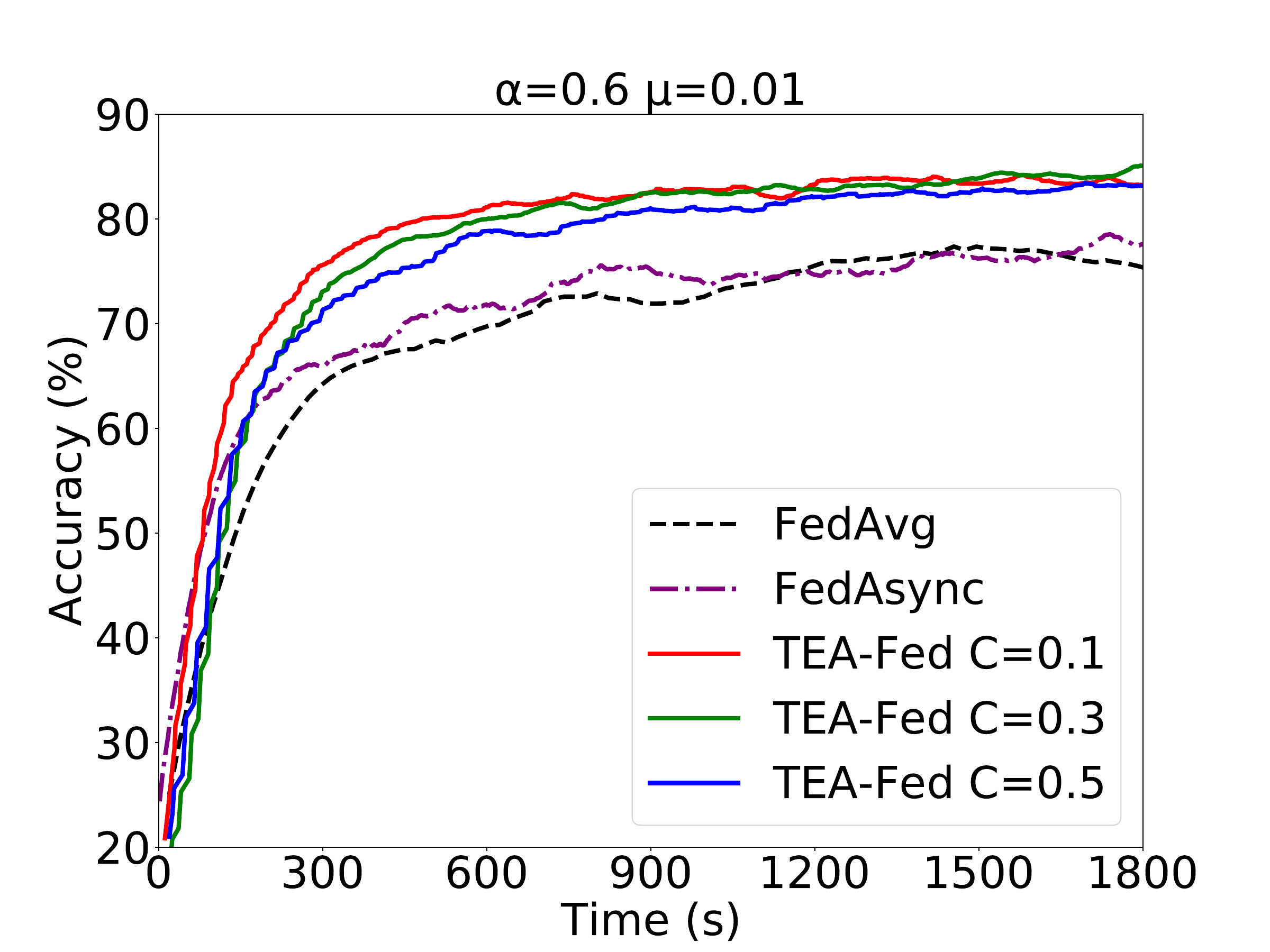}
\label{fig1a}}
\subfigure[IID]{\includegraphics[width=\figwidth\linewidth]{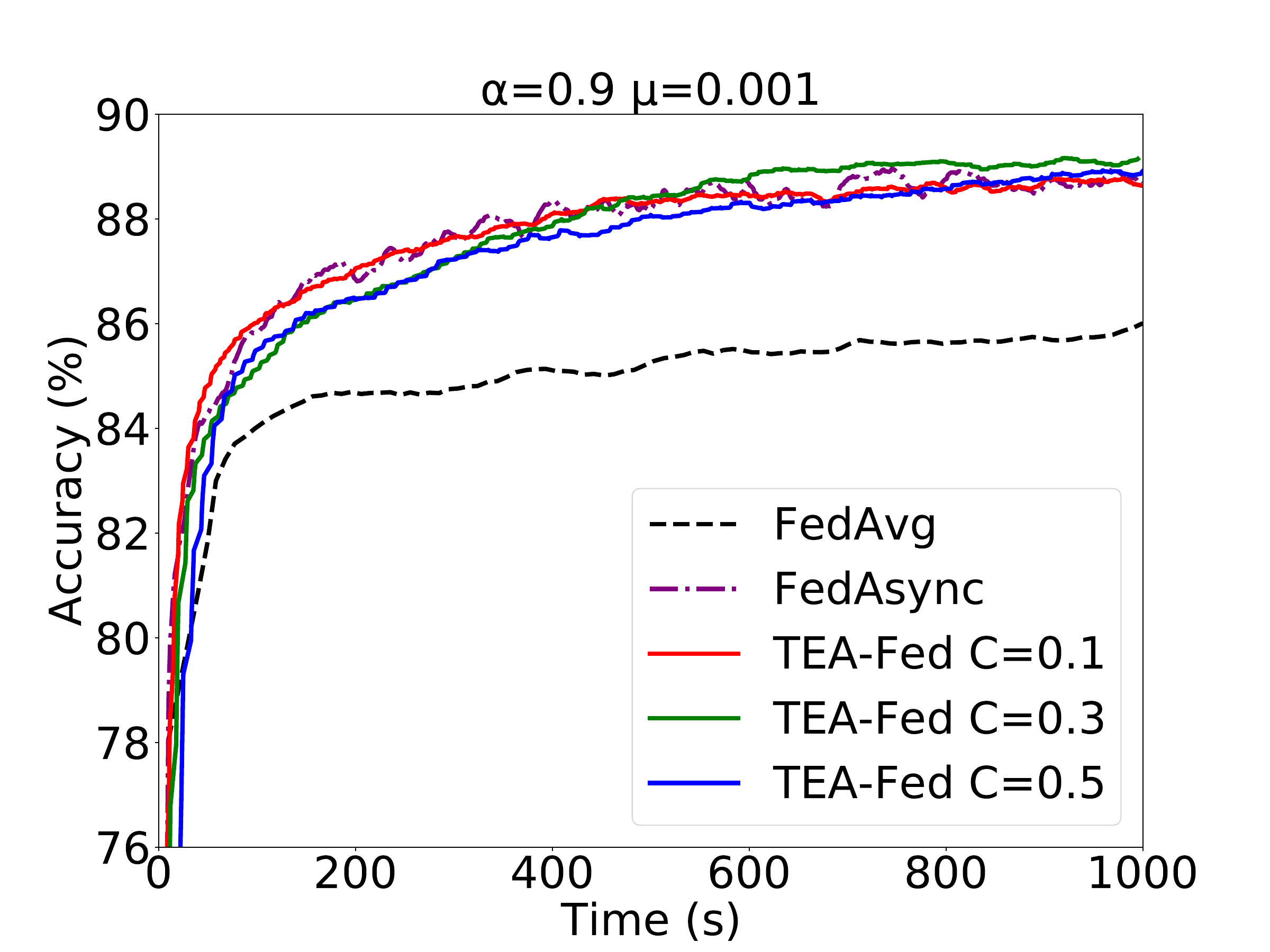}
\label{fig1b}}
\caption{Impacts of $C$ in terms of accuracy vs. training time with non-IID and IID dataset.}
\end{figure*}

We take FedAvg \cite{mcmahan2017communication} and FedASync \cite{xie2019asynchronous} as synchronous and asynchronous FL baselines. In the experiment, the maximum model staleness is maintained at 4 for FedASync, and 10 devices are randomly chosen for FedAvg in each cycle to carry out the local update. Additionally, we used the TEAStatic-Fed approach, in which the compression parameters are maintained constant during the training phase, to confirm the impact of dynamic parameter decay. We refer to TEASQ-Fed without sparsification or quantization by TEA-Fed in this part.

To verify the improvement of TEASQ-Fed in the wireless network environment relative to the baselines, we simulate a wireless IoT FL system with a Base Station (BS), i.e., server, and $M$ devices distributed in a circular area. The radius of the circular area is $R=600$ m or $1000$ m. The server is located at the center of the circular area. All devices are uniformly distributed in the circular area whose locations stay unchanged during the whole training process. The wireless bandwidth is $B = 20$ MHz. The path loss exponent is $\alpha = 3.76$. The transmission power of the BS, i.e., server, is $P_0 = 20$ dBm, and the transmit power of all devices is $P_k = 10$ dBm. The power spectrum density of the additive Gaussian noise is $N_0$ = -114 dBm/MHz. Thus the maximum achievable transmission rate (bits/s) of BS sending global models can be written as $r^d_k = B \log_{2} (1 + \frac{P_0 h^{2}_{0, k}}{B N_0})$ and the maximum achievable transmission rate (bits/s) of devices uploading local models can be written as $r^u_k = B \log_{2} (1 + \frac{P_k h^{2}_{k, 0}}{B N_0})$, where $h_{i, j}$ represents the corresponding channel gain and ``0'' represents the BS \cite{shi2020joint}. 

\begin{figure*}[t]
\centering
\subfigure[Non-IID]{\includegraphics[width=\figwidth\linewidth]{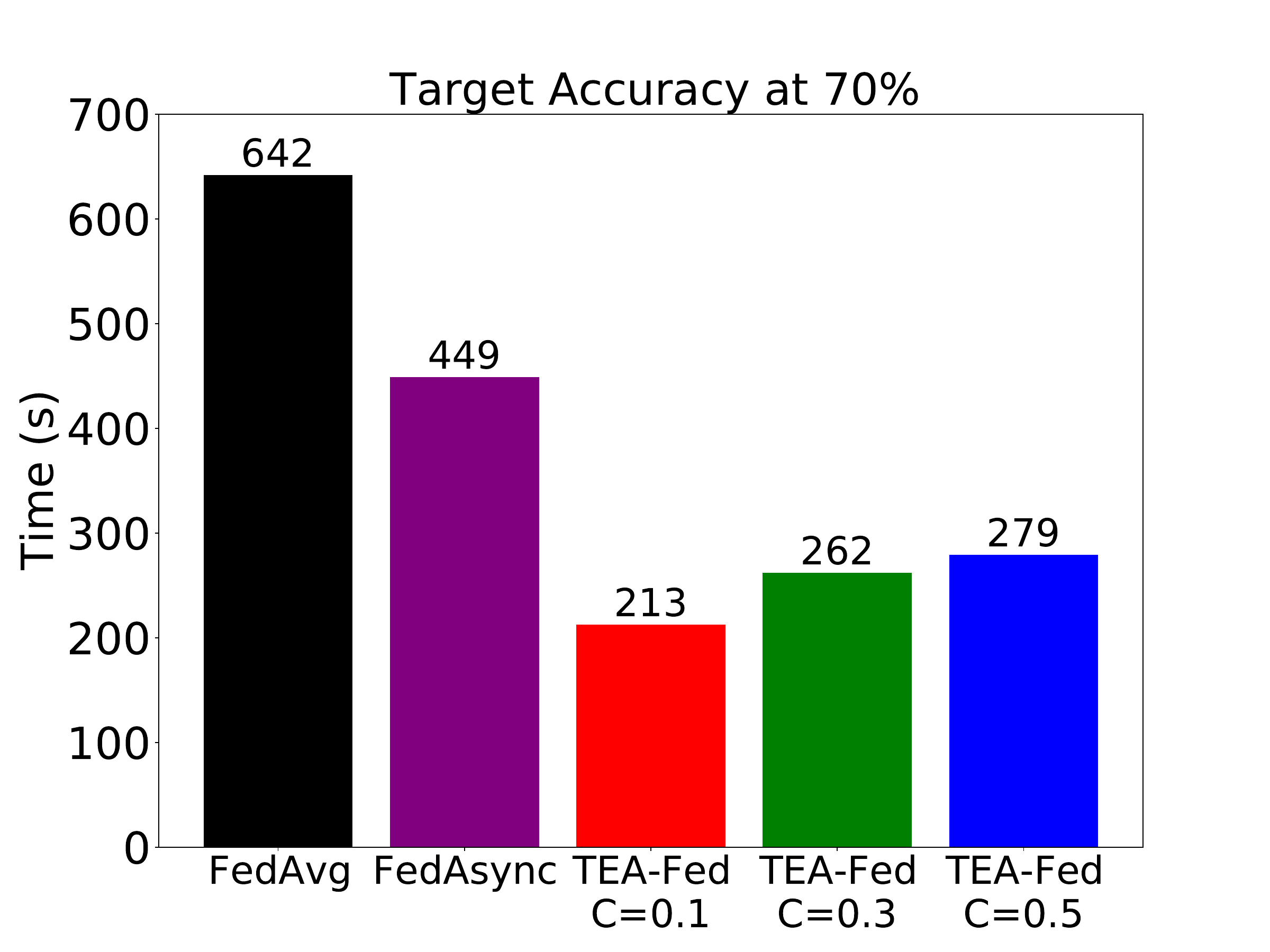}
\label{fig1c}}
\subfigure[IID]{\includegraphics[width=\figwidth\linewidth]{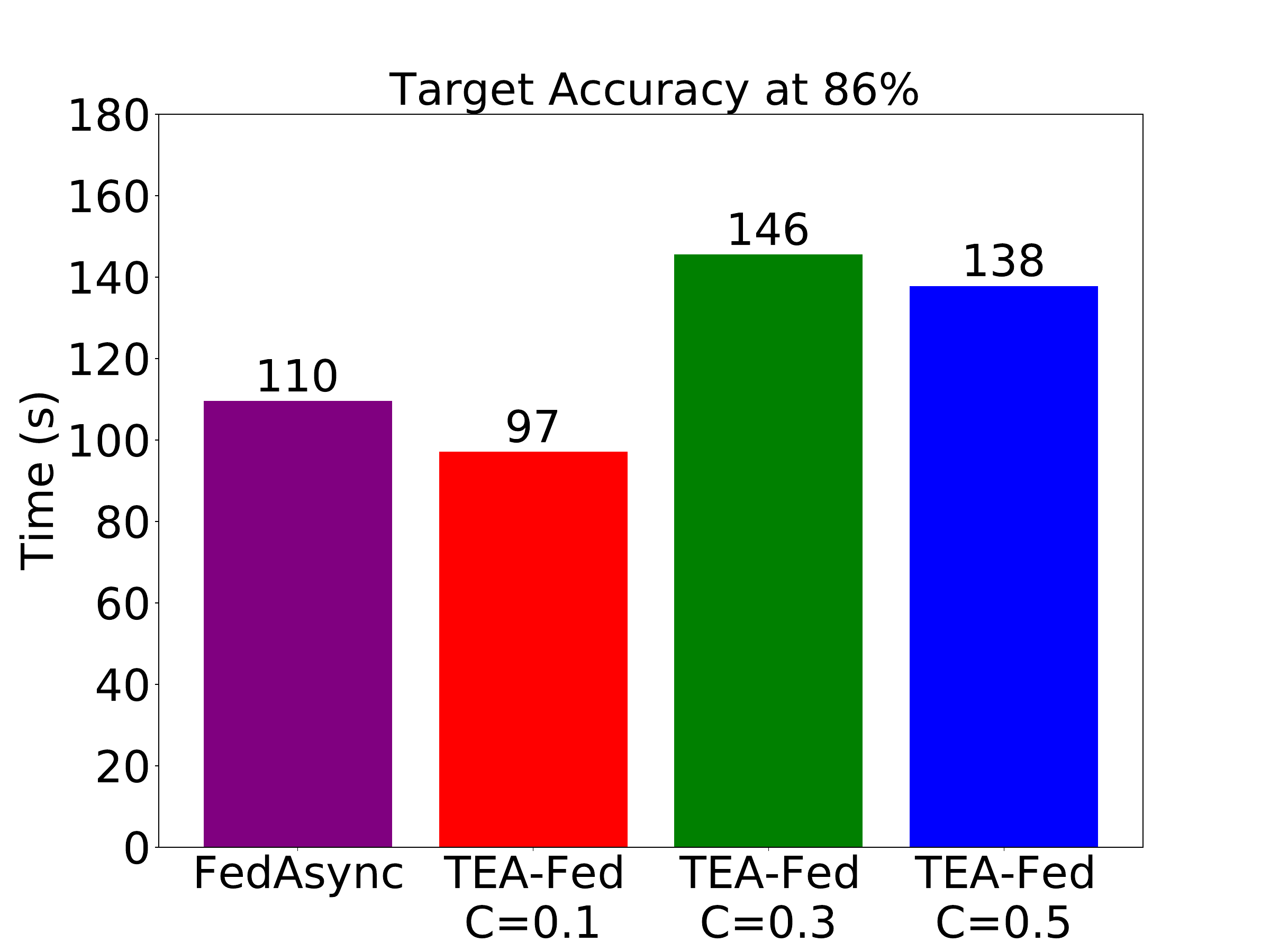}
\label{fig1d}}
\caption{Impacts of $C$ and time required to reach the target test accuracy with non-IID and IID dataset. In (b), FedAvg fails to reach the target accuracy.}
\end{figure*}

\subsection{Evaluation Results}
\label{subsection:results}

We carry out tests on a variety of hyper-parameters, such as $C$, $\alpha$, and $\mu$, and we examine how these hyper-parameters affect the convergence of the global model. In the meantime, we evaluate the model precision and convergence rates of several optimization strategies for various data distributions.

\subsubsection{Effects of $\boldsymbol{\mu}$}

We demonstrate the impact of various $\mu$ values on TEASQ-Fed without data compression (i.e., TEA-Fed)  using non-IID data in Figure~\ref{fig4}. The regularization weight parameter in the local optimization is non-negative, i.e., $\mu\ge0$. The weights of local updates with the same number of local epochs vary significantly because of the imbalanced data distribution\cite{zhao2018federated}. By exploiting a regularization penalty term, the local update limits the local update to be closer to the global model and allows the global model to converge more rapidly and smoothly \cite{li2020federated}. We provide a finite set of values for $\mu$, which are $\left\{0, 0.001, 0.005, 0.01, 0.1\right\}$. The experimental findings demonstrate that the convergence efficiency of the proposed algorithm with heterogeneous data can be significantly increased when $\mu > 0$. The corresponding accuracy, stability with non-IID data, and convergence speed of the global model can all be improved with appropriate $\mu$.

\begin{figure*}[!t]
\centering
\subfigure[Non-IID]{\includegraphics[width=\figwidth\linewidth]{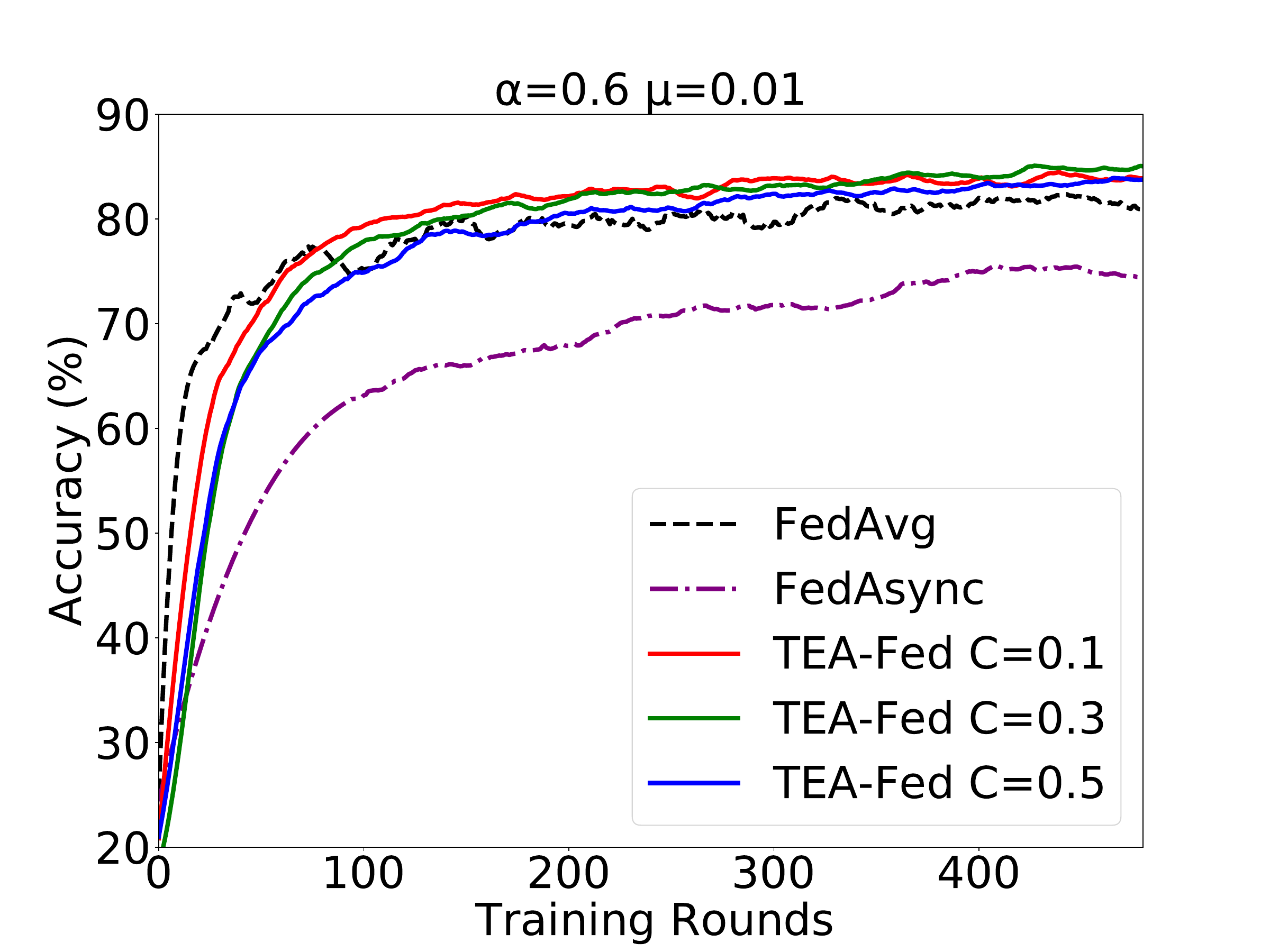}
\label{fig2a}}
\subfigure[IID]{\includegraphics[width=\figwidth\linewidth]{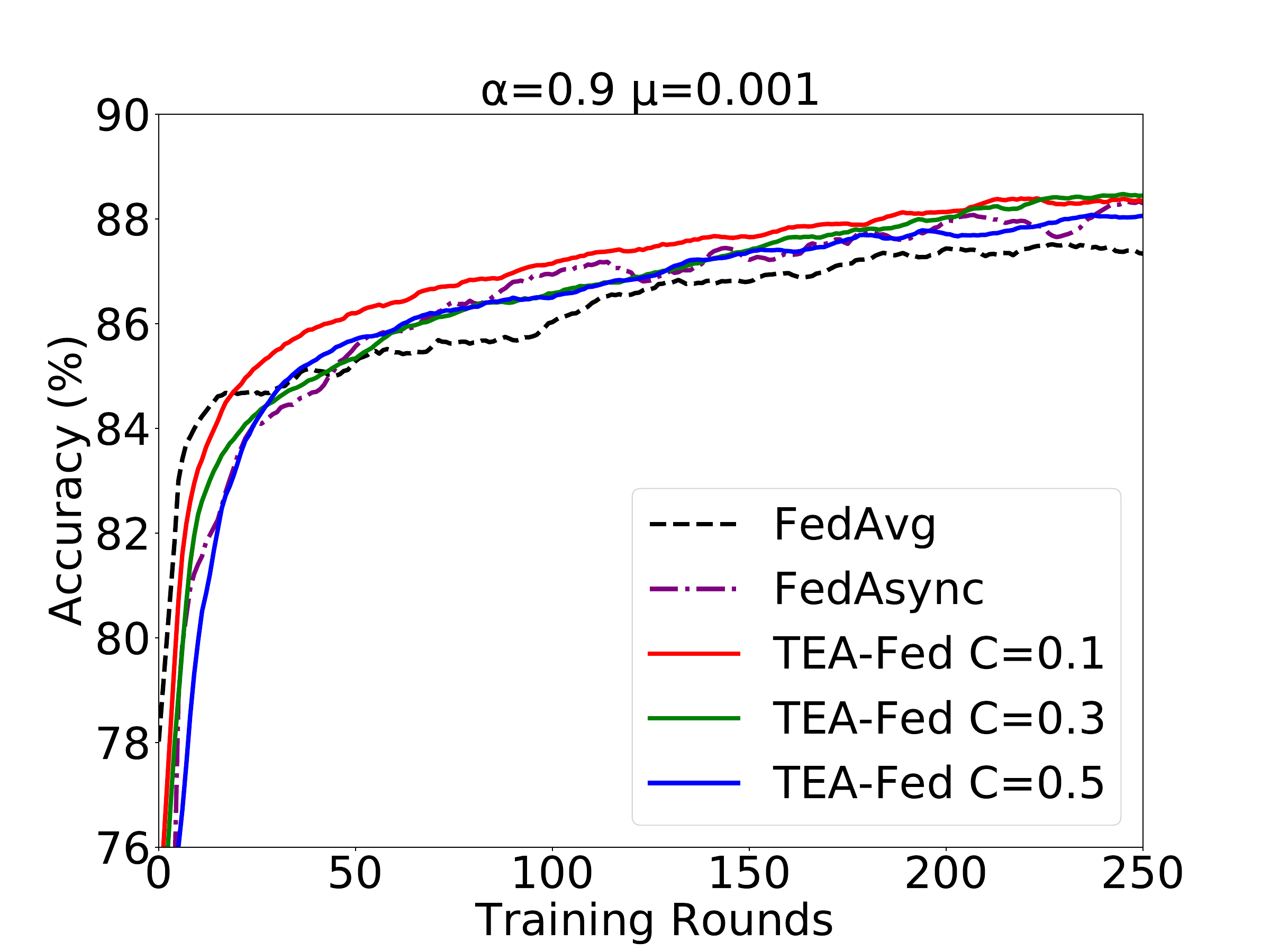}
\label{fig2b}}
\caption{Impacts of $C$ in terms of accuracy vs. training rounds with non-IID and IID dataset.}
\label{fig2}
\end{figure*}

\begin{figure*}[!t]
\centering
\subfigure[Non-IID]{\includegraphics[width=\figwidth\linewidth]{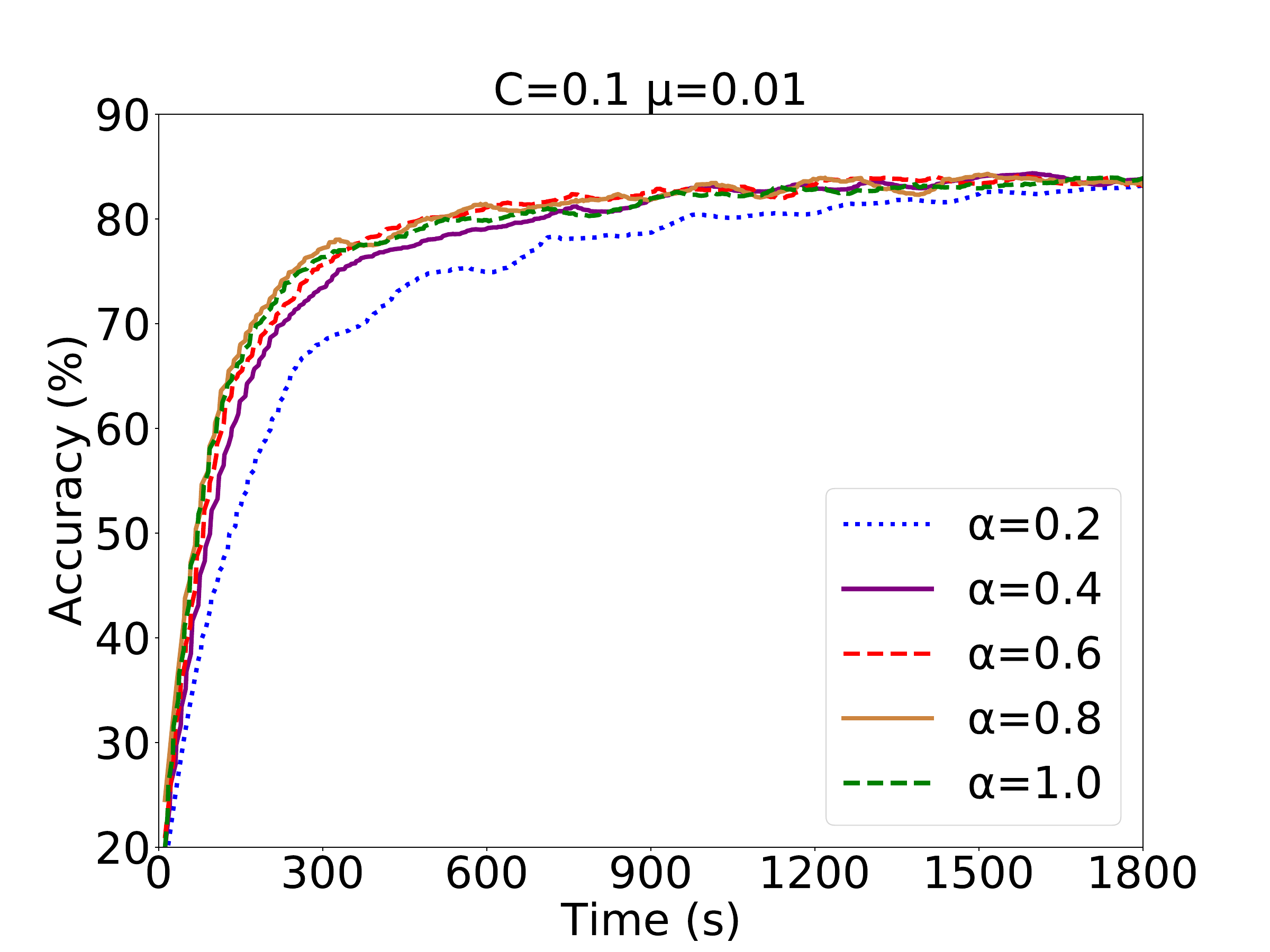}
\label{fig3a}}
\subfigure[IID]{\includegraphics[width=\figwidth\linewidth]{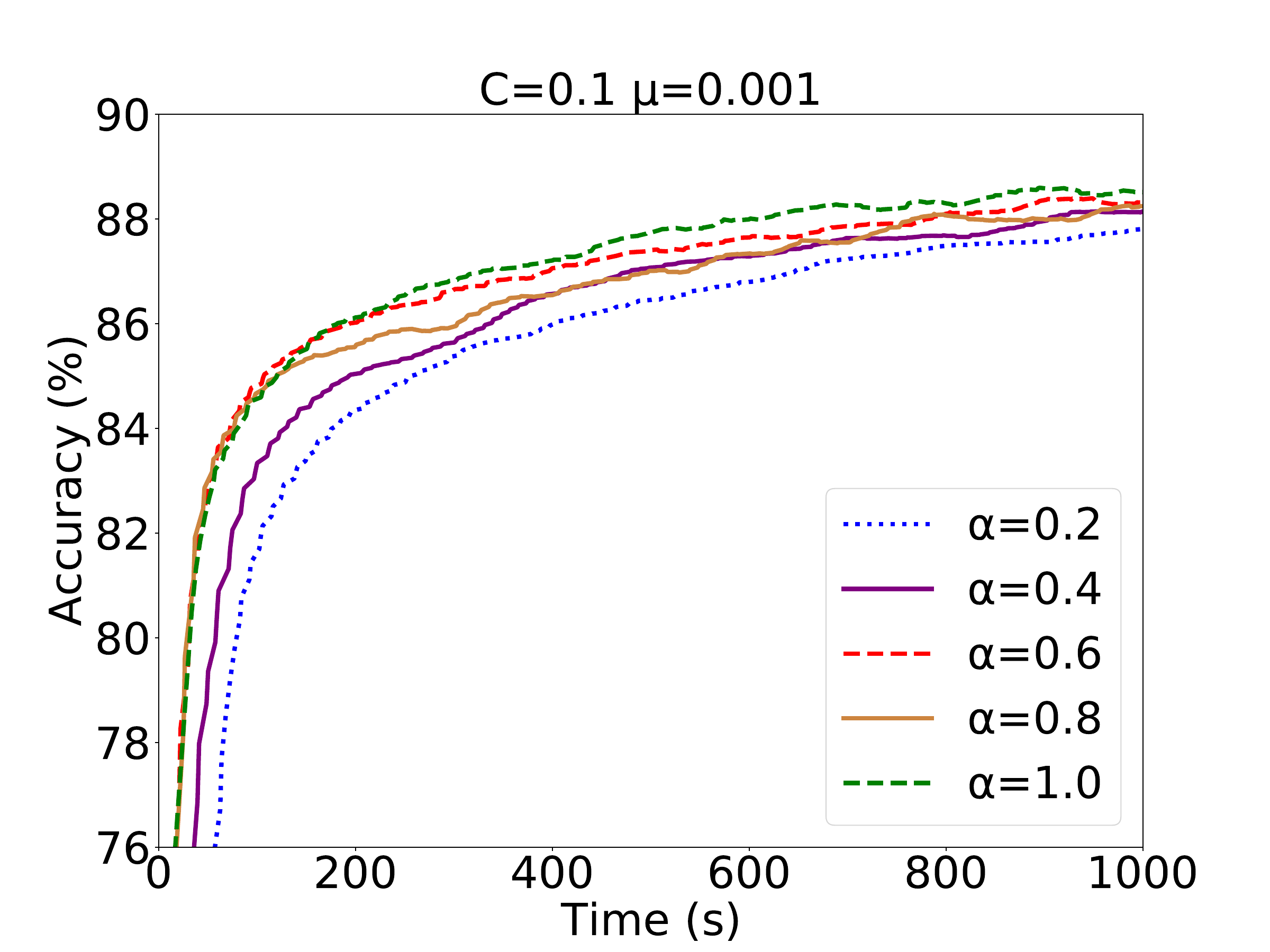}
\label{fig3b}}
\caption{Impacts of $\alpha$ in terms of accuracy vs. training time with non-IID and IID dataset.}
\label{fig3}
\end{figure*}

\subsubsection{Effects of $\boldsymbol{C}$} 

We demonstrate how TEASQ-Fed without data compression (i.e., TEA-Fed) converges with various $C$ values over time under non-IID and IID data distributions in Figures~\ref{fig1a} and \ref{fig1b}. 
The results show that the model accuracy and convergence speed of TEA-Fed are not directly proportional to the value of $C$.
Although increasing $C$ enhances the parallelism, too many devices are involved in the training process of the same epoch simultaneously, which could result in a significant delay. 
As a result, the performance rate is the highest when $C$ is 0.1, and TEA-Fed has a higher convergence efficiency than FedAvg and FedAsync. In addition, as depicted in Figures~\ref{fig1c} and ~\ref{fig1d}, TEA-Fed corresponds to a shorter time to obtain the desired precision. The training time required for each round is shorter than the baselines because it effectively decreases the waiting time. 
We show the convergence of TEA-Fed in terms of training rounds in Figure~\ref{fig2}. Evidently, the accuracy of TEA-Fed is significantly higher than that of baselines, further demonstrating its effectiveness under non-IID and IID data distributions.

\subsubsection{Effects of $\boldsymbol{\alpha}$}

The old global model and the new updated local models are aggregated with the hyper-parameter $\alpha$. The influence of the new update on the updated global model increases with the larger value of $\alpha$. Figure \ref{fig3} illustrates how TEA-Fed varies with various $\alpha$. As the influence of $\alpha$ between 0.4 and 0.9 on the convergence has little difference, the convergence of TEA-Fed does not depend on $\alpha$. Thus, $\alpha$ has limited influence on the convergence of TEA-Fed, which means TEA-Fed is robust to the change of $\alpha$.


\begin{figure*}[!t]
\centering
\subfigure[Non-IID]{\includegraphics[width=\figwidth\linewidth]{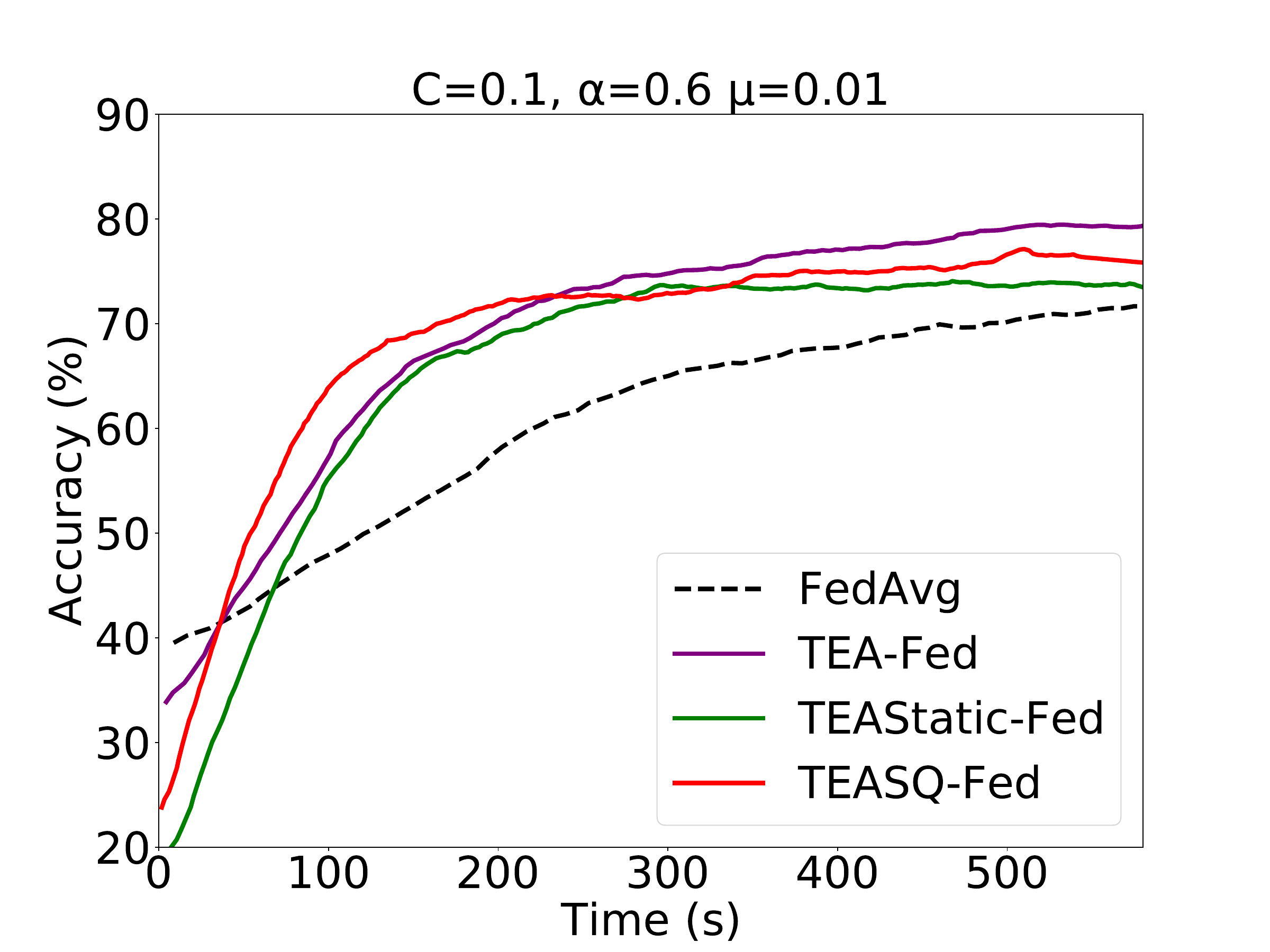}
\label{fig:compressnoniid}}
\subfigure[IID]{\includegraphics[width=\figwidth\linewidth]{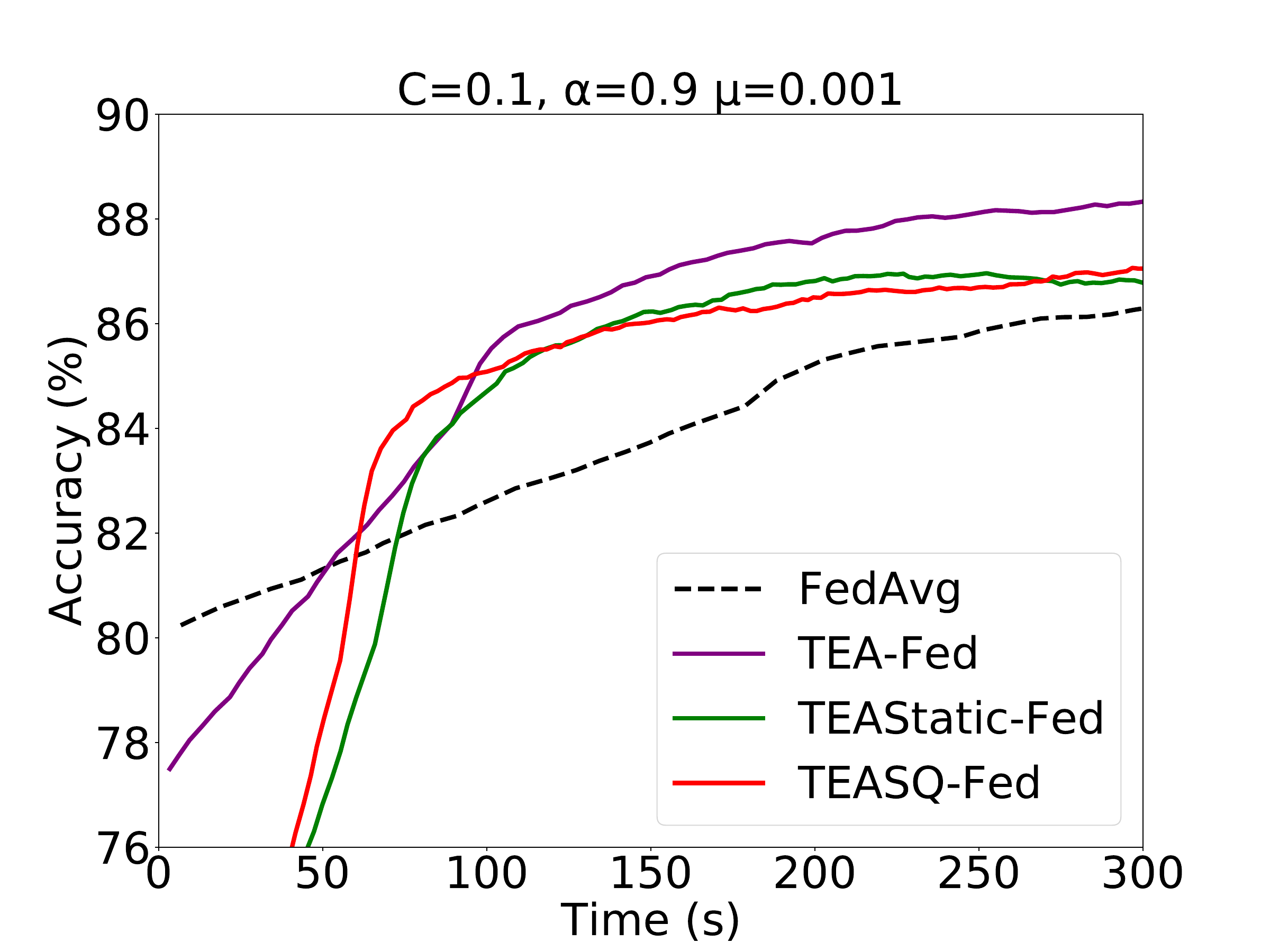}
\label{fig:compressiid}}
\caption{Impacts of compression in terms of accuracy vs. training time with non-IID and IID dataset.}
\label{figcompress}
\end{figure*}

\begin{table*}[!t]
\centering
\caption{The highest test accuracy within a given time budget using IID data.}
\label{tab:acciid}
\begin{tabular}{|c|c|c|c|c|c|c|c|c|}
\hline
\diagbox{Method}{Accuracy }{Time budget}
                        & 50s          & 60s            & 70s            & 80s            & 90s          & 100s           & 200s          & 300s            \\  
\hline
    FedAvg (IID)          & \bf{81.11\%} & 81.29\%        & 81.46\%        & 81.81\%        & 81.98\%      & 82.33\%        & 85.12\%       & 86.34\%         \\
\hline
    TEA-Fed (IID)         & 80.79\%      & \bf{81.34\%}   & 82.16\%        & 82.99\%        & 83.81\%      & \bf{85.23\%}   & \bf{87.64\%}  & \bf{88.32\%}    \\
\hline 
    TEAStatic-Fed (IID)   & 75.79\%      & 77.84\%        & 79.88\%        & 83.45\%        & 84.29\%      & 84.73\%        & 86.82\%       & 86.96\%         \\
\hline
    TEASQ-Fed (IID)   & 77.37\%      & 80.75\%        & \bf{83.96\%}   & \bf{84.65\%}   & \bf{84.96\%} & 85.08\%        & 86.47\%       & 87.09\%         \\
\hline
\end{tabular}
\end{table*}

\subsubsection{Effects of Compression} \label{p:compress} 

TEASQ-Fed exploits the compression parameters generated from the dynamic decay of $p_s$ and $p_q$ as shown in Algorithm~\ref{alg:search}. TEAStatic-Fed represents utilizing the parameters generated from Lines~\ref{al:searchbegin}-\ref{al:searchend} in Algorithm~\ref{alg:search}, where $p_s$ and $p_q$ are constant during the training process.
In Figure~\ref{fig:compressnoniid} and Figure~\ref{fig:compressiid}, we compare the accuracy corresponding to TEAStatic-Fed, TEASQ-Fed, and FedAvg with IID (Tables~\ref{tab:acciid}, \ref{tab:timeiid}) and non-IID (Tables~\ref{tab:accnoniid}, \ref{tab:timenoniid}) data, respectively. We can see that both TEAStatic-Fed and TEASQ-Fed have a faster convergence speed than FedAvg. TEA-Fed achieves a higher accuracy (up to 16.67\% as shown in Figure~\ref{fig2a}) than FedAvg and converges faster (up to twice faster as shown in Figure~\ref{fig1c} when target accuracy at 70\%) than FedAvg.
However, compared to TEA-Fed, due to the lossy data compression of data compression, the training of TEAStatic-Fed and TEASQ-Fed is more efficient at the early stage, while TEAStatic-Fed and TEASQ-Fed cannot converge to the same accuracy as TEA-Fed. When we can have a tight training time for a modest accuracy requirement, dynamic data compression (TEASQ-Fed) can achieve a higher accuracy (up to 10.08\% compared with TEA-Fed and up to 32.40\% compared with FedAvg as shown in Table~\ref{tab:accnoniid}) within a short training time (100s) or a shorter training time (up to 45.73\% compared with TEA-Fed and twice faster compared with FedAvg as shown in Table~\ref{tab:timenoniid}) with a modest target accuracy (68\%).
In addition, the required storage space corresponding to the dynamic data compression with sparsification and quantization is 44.07\% (when uploading local
models) smaller than that of FedAvg, as shown in Table~\ref{tab:space}. 

\begin{figure}[!t]
    \centering
    \includegraphics[width=\figwidth\linewidth]{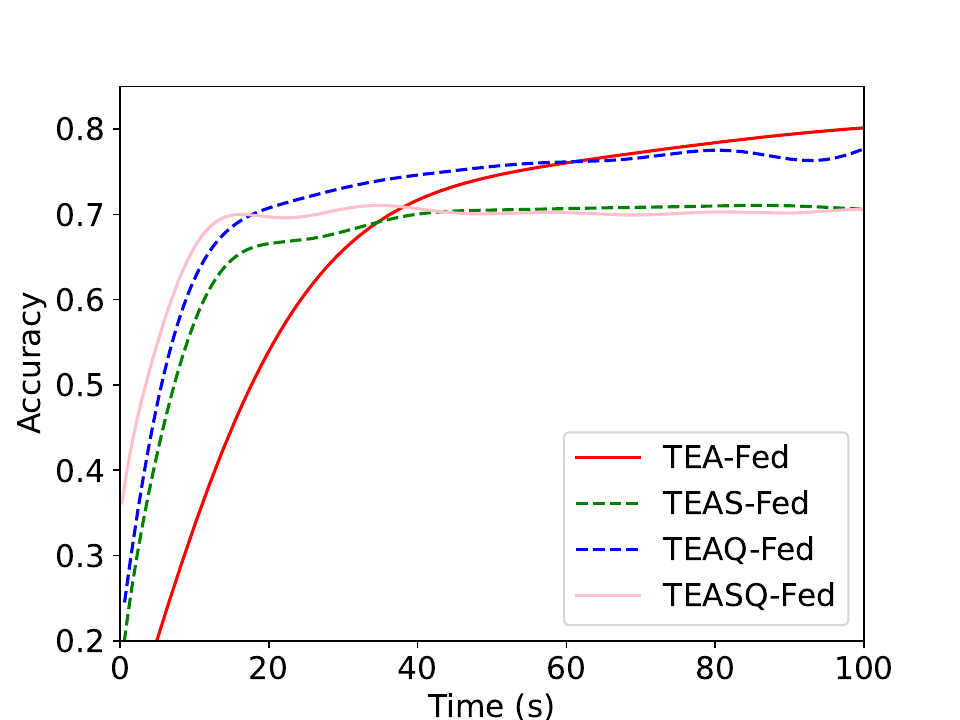}
    \caption{\rev{Ablation experiments of compression methods.}}
    \label{fig:ablation}
\end{figure}

\rev{We further conduct ablation experiments on the used compression methods. In addition to the TEA-Fed (i.e., TEASQ-Fed without data compression), we evaluate the performance of our algorithm with only one single compression method, which is referred to as TEAS-Fed (i.e., TEA-Fed with sparsification) and TEAQ-Fed (i.e., TEA-Fed with quantization). In Figure \ref{fig:ablation}, we show the accuracy of TEA-Fed, TEAS-Fed, TEAQ-Fed and TEASQ-Fed during the training progresses. According to the result, both TEAS-Fed and TEAQ-Fed with a single compression method can achieve faster training than TEA-Fed, and TEASQ-Fed which combines them can accelerate the training process more significantly. However, the performance of all methods with compression suffer from some degree of degradation, which is the cost of using compression.}

\begin{table*}[!t]
\centering
\caption{Time consumed to reach the target accuracy with IID data. ``-'' represents that the target is not achieved.}
\label{tab:timeiid}
\begin{tabular}{|c|c|c|c|c|c|c|c|c|}
\hline
\diagbox{Method}{Time }{Target Accuracy}
                        & 81\%         & 82\%           & 83\%         & 84\%          & 85\%         & 86\%           & 87\%           & 88\%              \\ 
\hline
    FedAvg (IID)          & \bf{45.26s}  & 97.16s        & 134.01s       & 176.51s       & 205.15s      & 262.94s        & -              & -                 \\
\hline
    TEA-Fed (IID)         & 55.73s       & 70.63s        & 84.82s        & 94.98s        & 102.58       & \bf{116.37s}   & \bf{154.63s}   & \bf{233.47s}      \\
\hline
    TEAStatic-Fed (IID)   & 75.63s       & 77.61s        & 82.59s        & 88.26s        & 104.32s      & 138.76s        & -              & -                 \\
\hline
    TEASQ-Fed (IID)       & 63.48s       & \bf{65.17s}   & \bf{67.5s}    & \bf{74.29s}   & \bf{95.3s}   & 147.71s        & 290.28s        & -                 \\
\hline
\end{tabular}
\end{table*}

\begin{table*}[!t]
\centering
\caption{The highest test accuracy within a given time budget using non-IID data.}
\label{tab:accnoniid}
\begin{tabular}{|c|c|c|c|c|c|c|c|c|}
\hline
\diagbox{Method}{Accuracy }{Time budget}
                            & 50s            & 100s           & 125s           & 150s           & 175s         & 200s           & 400s           & 600s       \\  
\hline
    FedAvg (non-IID)          & 42.99\%        & 47.84\%        & 49.92\%        & 52.69\%        & 54.77\%      & 58.22\%        & 67.67\%        & 71.66\%    \\
\hline
    TEA-Fed (non-IID)         & 44.69\%        & 57.54\%        & 62.43\%        & 65.91\%        & 67.95\%      & 70.01\%        & \bf{76.90\%}   & \bf{79.52\%}  \\
\hline
    TEAStatic-Fed (non-IID)   & 37.33\%        & 54.44\%        & 60.45\%        & 64.84\%        & 67.03\%      & 68.31\%        & 73.73\%        & 74.07\%     \\
\hline
    TEASQ-Fed (non-IID)       & \bf{47.94\%}   & \bf{63.34\%}   & \bf{66.85\%}   & \bf{68.61\%}   & \bf{70.26\%} & \bf{71.67\%}   & 75.04\%        & 77.12\%       \\
\hline
\end{tabular}
\end{table*}

\begin{table*}[!t]
\centering
\caption{Time consumed to reach the target accuracy with non-IID data.}
\label{tab:timenoniid}
\begin{tabular}{|c|c|c|c|c|c|c|c|c|}
\hline
\diagbox{Method}{Time }{Target Accuracy}
                         & 68\%         &    69\%           & 70\%           & 71\%           & 72\%         & 73\%           & 75\%          & 79\%            \\  
\hline
    FedAvg (non-IID)       & 402.71s      &    441.44s        & 483.52s        & 548.49s        & -            & -              & -             & -               \\
\hline
    TEA-Fed (non-IID)       & 193.47s      &    203.23s        & 224.49s        & 204.86s        & 220.72s      & \bf{238.23s}   & \bf{307.11s}  & \bf{507.74s}    \\
\hline
    TEAStatic-Fed (non-IID)    & 198.93s      &    208.6s         & 229.83s        & 238.23s        & 263.83s      & 288.92s        & -             & -               \\
\hline
    TEASQ-Fed (non-IID)     & \bf{132.76s} &    \bf{150.10s}   & \bf{166.58s}   & \bf{202.43s}   & \bf{206.56s} & 306.27s        & 370.56s       & -               \\
\hline
\end{tabular}
\end{table*}

\rev{
\subsubsection{Comparison with SOTAs}

\begin{figure}[!t]
    \centering
    \includegraphics[width=\figwidth\linewidth]{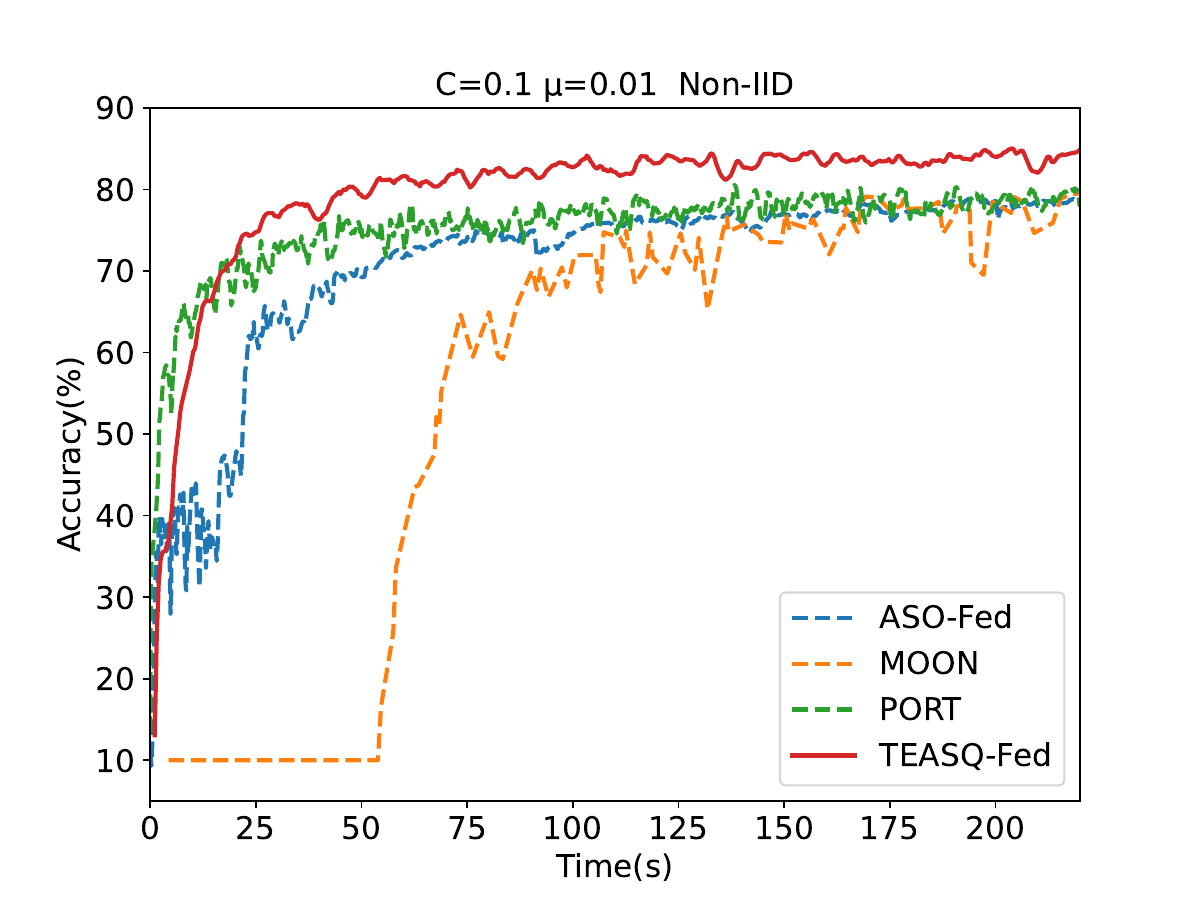}
    \caption{\rev{Comparison with asynchronous methods ASO-Fed, PORT and synchronous method MOON.}}
    \label{fig:compare sota}
\end{figure}

We compare our TEASQ-Fed with state-of-the-art baselines: asynchronous methods PORT \cite{Su2022How} and ASO-Fed \cite{chen2020asynchronous} as well as a synchronous method MOON \cite{li2021model}. As shown in Figure \ref{fig:compare sota}, TEASQ-Fed outperforms the other baselines in terms of the model accuracy, while it also has quicker convergence speed than most baselines except for PORT. Therefore, TEASQ-Fed has both advantages of accuracy and speed.
}

\subsection{Discussion}

Extensive experiments with diverse data distributions demonstrate that TEASQ-Fed has excellent performance for FL tasks and has significantly superior overall convergence than FedAvg and FedAsync. We summarize the improvements of TEASQ-Fed in the following three aspects:

\begin{itemize}

\item Faster convergence during model training. The asynchronous training mechanism makes it possible for more idle devices to participate in the training process, hastening the convergence of the model. The convergence speed of TEASQ-Fed is faster than the baseline methods, even if the $C$ value is larger.

\item Higher convergence accuracy. We assign diverse importance based on the staleness of the updated local models during the model aggregation process. The rebalance of models improves the convergence rate and improves accuracy by reducing the impact of the staleness of local updates on the global model. 

\item Robust with heterogeneous data. TEASQ-Fed can achieve high convergence accuracy with highly heterogeneous data. With the regular penalty term within the local optimization on unbalanced data, we prevent divergence and increase the stability of the convergence of the global model. When dealing with non-IID data, the adjustment of the regularization value $\mu$ can lead to excellent performance in terms of both accuracy and convergence rate.

\end{itemize}

\begin{table}[!t]
\centering
\caption{The maximum storage space required during training.}
\label{tab:space}
\begin{tabular}{|c|c|c|}
\hline
\diagbox[width=11em]{Method}{Size}{Type}
                        & Global model                   & Local models \\ 
\hline
FedAvg (IID)              & 794.66KB                       & 794.66KB    \\
\hline
TEA-Fed (IID)             & 878.41KB                       & 878.41KB    \\
\hline
TEAStatic-Fed (IID)       & 575.70KB                       & 496.53KB    \\
\hline
TEASQ-Fed (IID)           & 529.77KB                       & 498.92KB    \\  
\hline
FedAvg (non-IID)          & 794.66KB                       & 794.66KB    \\
\hline
TEA-Fed (non-IID)         & 878.41KB                       & 878.41KB    \\
\hline
TEAStatic-Fed (non-IID)   & 519.78KB                       & 437.09KB    \\
\hline
TEASQ-Fed (non-IID)       & 470.93KB                       & 444.43KB    \\  
\hline
\end{tabular}
\end{table}

\section{Conclusion}
\label{sec:conclusion}

In this paper, we propose TEASQ-Fed, an innovative asynchronous federated technique with sparsification and quantization that intends to efficiently carry out FL training in edge computing. Through asynchronous training, TEASQ-Fed fully utilizes the downtime of edge devices, allowing more devices than with conventional methods to participate in the training process. In the meantime, TEASQ-Fed can significantly reduce the training time for the training process of FL and successfully minimize synchronization overhead. The extensive experimentation demonstrates that our approach has significant advantages with heterogeneous data in terms of accuracy (up to 16.67\%) and training time (up to twice faster). The future scope of this work includes: (1) The investigation of the potential parallel training and transmission of different versions of models to further improve the efficiency of asynchronous FL; (2) The testbed experiment of the proposed algorithms to justify the performance in the real world.


\section*{Acknowledgments}
This work was partially (for J. Jia) supported by the Collaborative Innovation Center of Novel Software Technology and Industrialization, Project Funded by the Priority Academic Program Development of Jiangsu Higher Education Institutions, Suzhou Frontier Science and Technology Program (Project SYG202310). 
Parts of the experiments in this paper were carried out on the Baidu Data Federation Platform (Baidu FedCube). For usages, please contact us via \{fedcube, shubang\}@baidu.com.


\bibliography{reference}

\end{document}